\documentclass{article} 
\usepackage{iclr2025_conference,times}

\usepackage{amsmath,amsfonts,bm}

\def\Tabref#1{Tab.~\ref{#1}}

\def\Figref#1{Fig.~\ref{#1}}

\def\Secref#1{Section~\ref{#1}}

\def\Eqref#1{Eq.~\eqref{#1}}

\def\Algref#1{Alg.~\ref{#1}}

\def\1{\bm{1}}

\def\ru{{\textnormal{u}}}
\def\rv{{\textnormal{v}}}

\def\rvepsilon{{\mathbf{\epsilon}}}

\def\rvg{{\mathbf{g}}}

\def\rvp{{\mathbf{p}}}

\def\rvx{{\mathbf{x}}}
\def\rvy{{\mathbf{y}}}
\def\rvz{{\mathbf{z}}}

\def\rmV{{\mathbf{V}}}

\def\vy{{\bm{y}}}

\def\mI{{\bm{I}}}

\def\mV{{\bm{V}}}

\DeclareMathAlphabet{\mathsfit}{\encodingdefault}{\sfdefault}{m}{sl}
\SetMathAlphabet{\mathsfit}{bold}{\encodingdefault}{\sfdefault}{bx}{n}

\usepackage{enumitem}
\usepackage{hyperref}
\usepackage{url}
\usepackage{algpseudocode}
\usepackage{algorithm}
\usepackage{booktabs} 
\usepackage{graphicx}
\usepackage{amssymb}
\usepackage{xspace}
\usepackage{wrapfig}
\usepackage{makecell}
\usepackage{listings}
\usepackage{xcolor}

\definecolor{codegreen}{rgb}{0,0.6,0}
\definecolor{codegray}{rgb}{0.5,0.5,0.5}
\definecolor{codepurple}{rgb}{0.58,0,0.82}
\definecolor{backcolour}{rgb}{0.95,0.95,0.92}

\lstdefinestyle{mystyle}{
    backgroundcolor=\color{backcolour},   
    commentstyle=\color{codegreen},
    keywordstyle=\color{magenta},
    numberstyle=\tiny\color{codegray},
    stringstyle=\color{codepurple},
    basicstyle=\ttfamily\footnotesize,
    breakatwhitespace=false,         
    breaklines=true,                 
    captionpos=t,                    
    keepspaces=true,                 
    numbers=left,                    
    numbersep=5pt,                  
    showspaces=false,                
    showstringspaces=false,
    showtabs=false,                  
    tabsize=2
}

\newcommand{\rebuttal}[1]{#1}

\newcommand{\ours}{\textsc{CryoFM}\xspace}

\newcommand{\fV}{\tilde{\rmV}}
\newcommand{\fy}{\tilde{\rvy}}

\newcommand{\nnvtxty}{\rebuttal{\rv_\Theta(t,\rvx_t|\rvy)}}
\newcommand{\nnvtxt}{\rebuttal{\rv_\Theta(t,\rvx_t)}}
\newcommand{\nnv}{\rebuttal{\rv_\Theta}}

\allowdisplaybreaks

\usepackage{caption}
\usepackage[compact]{titlesec}
\usepackage{sidecap}
\newcommand{\beforesec}{0.1cm}
\newcommand{\aftersec}{0.1cm}
\titlespacing*{\section}{0pt}{\beforesec}{\aftersec}
\titlespacing*{\subsection}{0pt}{\beforesec}{\aftersec}
\titlespacing*{\subsubsection}{0pt}{\beforesec}{\aftersec}

\title{CryoFM: A Flow-based Foundation Model for Cryo-EM Densities}

\author{Yi Zhou\thanks{Equal Contribution, $^\dagger$ Corresponding Author}~~,~~~~Yilai Li$^*$,~~~~Jing Yuan$^*$,~~~~Quanquan Gu$^\dagger$\\
Bytedance Research\\ 
\texttt{\{zhouyi.naive,yilai.li,yuanjing.eugene,quanquan.gu\}@bytedance.com} \\
}
\iclrfinalcopy 
\begin{document}

\maketitle
\vspace{-0.5cm}
\begin{abstract}


Cryo-electron microscopy (cryo-EM) is a powerful technique in structural biology and drug discovery, enabling the study of biomolecules at high resolution. Significant advancements by structural biologists using cryo-EM have led to the production of \rebuttal{around 40k} protein density maps at various resolutions \rebuttal{on EMDB}\footnote{EMDB~\citep{emdb} is a public repository for cryo-EM density volumes.}. However, cryo-EM data processing algorithms have yet to fully benefit from our knowledge of biomolecular density maps, with only a few recent models being data-driven but limited to specific tasks. In this study, we present \ours, a foundation model designed as a generative model, learning the distribution of high-quality density maps and generalizing effectively to downstream tasks. Built on flow matching, \ours is trained to accurately capture the prior distribution of biomolecular density maps. Furthermore, we introduce a flow posterior sampling method that leverages \ours as a flexible prior for several downstream tasks in cryo-EM and cryo-electron tomography (cryo-ET) without the need for fine-tuning, achieving state-of-the-art performance on most tasks and demonstrating its potential as a foundational model for broader applications in these fields.

\end{abstract}

\section{Introduction}

Cryo-electron microscopy is an important technique in structural biology and drug discovery, allowing the determination of high-resolution 3D structures of biomolecules that are difficult to study through conventional methods, offering insights into molecular mechanisms and aiding in fields like drug discovery \citep{nogales2015cryo}. A major component of cryo-EM data processing involves reconstructing 3D structures from noisy 2D projections of particles. It can be formulated as an inverse problem \citep{bendory2020single,singer2020computational}, which aims at recovering the signal $\rvx \in \mathbb{R}^{n}$ (i.e., a clean protein density) from the observation $\rvy \in \mathbb{R}^{m}$ \footnote{The signal $\rvx \in \mathbb{R}^{D\times D\times D}$ denotes a clean protein density, where $D$ is the side length. The observation $\rvy$ can have any arbitary shape, so we use $\rvx \in \mathbb{R}^{n}$ for ease of notation but without loss of generality.}. Under this formulation, following Bayesian statistics, the objective is to sample a density from the posterior $p(\rvx|\rvy)$, which can be factorized to $p(\rvx|\rvy)\propto p(\rvy|\rvx)p(\rvx)$. Thus, the prior distribution becomes essential for guiding the reconstruction process and improving the accuracy of the resulting structures.

\citet{scheres2012relion2} first introduced a Gaussian distribution as the prior $p(\rvx)$ in 3D reconstruction for cryo-EM, effectively functioning as a frequency-dependent low-pass filter, which has proven useful in many cases. Building on this, recent works have explored more sophisticated regularizers rather than explicitly defining a prior distribution \citep{punjani2020non, tegunov2021multi, kimanius2021exploiting, li2023cryostar, schwab2024dynamight, kimanius2024data, liu2024overcoming}. While these approaches were initially developed for 3D reconstruction tasks, their methodologies are closely aligned with the methods for density map modification and post-processing \citep{jakobi2017model,ramirez2020automatic,terwilliger2020cryo,kaur2021local, sanchez2021deepemhancer,he2023improvement}. Both share the common objective of improving the quality of cryo-EM maps, whether during refinement or in post-processing. Among all these approaches, a line of methods has emerged that utilizes data-driven models to directly learn $ p(\rvx|\rvy) $ from data, showing strong potential \citep{sanchez2021deepemhancer,he2023improvement,kimanius2024data}. While powerful, these models are often designed for specific tasks, limiting their general applicability and versatility. 

\begin{figure}[th]
\centering
\includegraphics[width=1.0\textwidth]{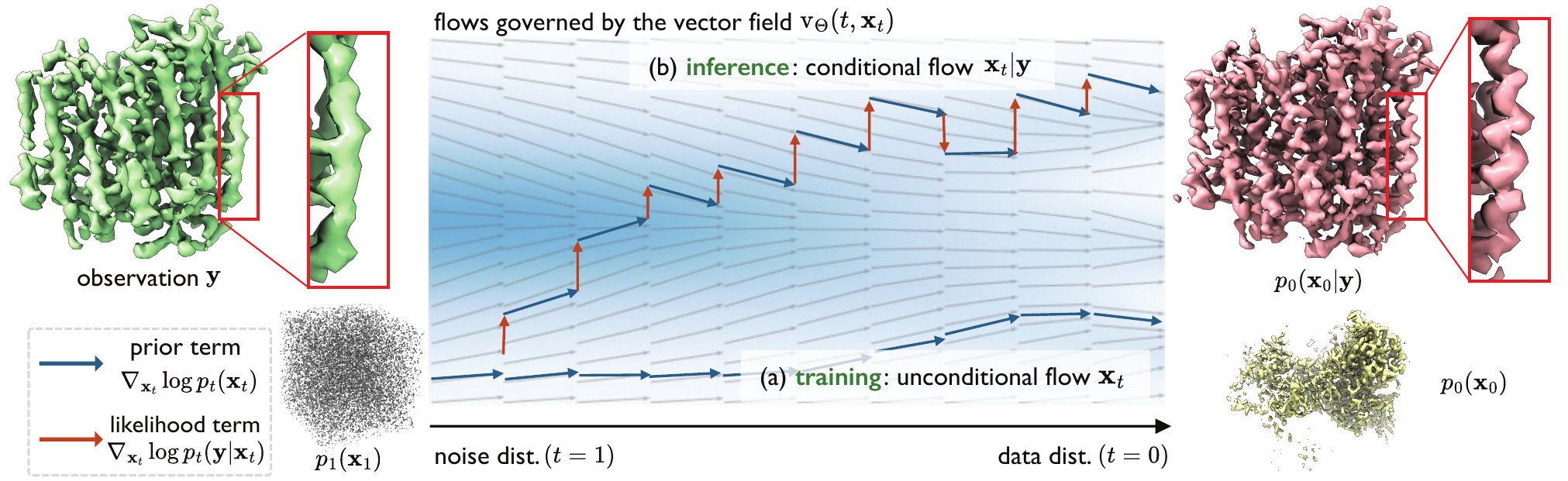}
\caption{The overview of \ours. In the training stage, \ours learns a vector field $\nnvtxt$, whose corresponding probability flow generates the data distribution $p_0(\rvx_0)$ of high-quality protein densities. In the inference stage, given an observation $\rvy$, a likelihood term $p_t(\rvy|\rvx_t)$ is incorporated to convert the unconditional vector field $\nnvtxt$ to a conditional one $\nnvtxty$, so that we can sample from the posterior distribution $p_0(\rvx_0|\rvy)$. This enables signal restoration of the density map, resulting in improved resolution of the alpha helices in the shown case.
}
\label{fig:intro}
\end{figure}




Foundation models leverage prior knowledge and provide versatility across a wide range of applications. In particular, they have revolutionized AI for protein structure prediction and design by utilizing vast datasets to tackle diverse tasks \citep{jumper2021highly, baek2021accurate, lin2023evolutionary, krishna2024generalized, hayes2024simulating, abramson2024accurate, wang2024diffusion}. \rebuttal{Chroma \citep{ingraham2023illuminating} and RFDiffusion \citep{watson2023novo} are two foundation models for protein structure modeling, which are broadly applicable to many structure-related tasks.} However, experimental electron density maps, a key modality in structural biology, have largely been neglected in this space. Meanwhile, the cryo-EM field has not yet taken full advantage of foundation models, which integrate prior knowledge and offer flexibility across multiple applications. \rebuttal{\citet{wang2024diffmodeler} developed a condition diffusion model in the density space, which achieves superior results by refining the map for model building, but is limited to single downstream task.} In this study, we present a \textit{foundation model}, \ours, that directly models the prior distribution $p(\rvx)$ by learning the distribution of high-quality cryo-EM density maps through flow matching. Notably, flow matching \citep{lipman2022flow,liu2022flow} is a well-established generative model, which achieves state-of-the-art results in many applications \citep{dao2023flow,esser2024scaling,yim2023fast}. The key ingredient of flow matching is a time-dependent vector field $\nnvtxt$, which can be employed to progressively denoise a noise vector into a data sample and is closely related to the score of the prior distribution $\nabla_{\rvx_t} \log p(\rvx_t)$. In \ours, when applied to downstream tasks, we draw a sample from the posterior distribution with its score \citep{ho2020denoising,song2021scorebased}: $\nabla_\rvx \log p(\rvx|\rvy) = \nabla_\rvx \log p(\rvx) + \nabla_\rvx \log p(\rvy|\rvx)$. In contrast to expert-designed prior, a data-driven prior is more powerful and expressive. Compared with models that directly learn the posterior, a decomposition of the prior and likelihood is more robust and versatile. As long as we can derive the likelihood term for a downstream task, we can combine it with the learned prior in a plug-and-play fashion \citep{zhang2021plug,venkatakrishnan2013plug} to sample from the posterior.


\ours consists of two models: \ours-S, which captures fine local details at high resolution, and \ours-L, which focuses on the overall global shape of the biomolecule at medium to low resolution\footnote{Unless stated otherwise, \ours refers to \ours-S throughout this paper.}. \Figref{fig:intro} provides an overview of \ours. In the training stage, we regress the vector field with high-quality density data downloaded from the EMDB database \citep{10.1093/nar/gkad1019}. When applied to downstream tasks, we propose a flow posterior sampling method to convert the vector field $\nnvtxt$ to a conditional one $\nnvtxty$, which \rebuttal{can be used to sample from} the posterior distribution $p(\rvx|\rvy)$. When applied to various downstream tasks, \ours achieves state-of-the-art results on most of the experiments, demonstrating the power and potential of \ours as a \textit{foundation model} for solving more complex problems in cryo-EM and cryo-ET.


Our main contributions are summarized as follows: (1) we present the first flow-based generative model as a foundation model that learns the distribution of high-quality cryo-EM density maps; (2) we derive a flow posterior sampling algorithm, enabling \ours to be effectively utilized as a prior for various downstream tasks; (3) in an unsupervised manner, we achieve better performance in several downstream tasks compared to DeepEMhancer \citep{sanchez2021deepemhancer}, EMReady \citep{he2023improvement} and spIsoNet \citep{liu2024overcoming}; (4) we explore different model architectures and configurations to optimize the training of generative models on large biomolecular density maps.


\section{Related Work}
In this section, we will briefly review the mostly related work on cryo-EM and diffusion/flow matching for inverse problems. A more detailed discussion on these topics can be found in Section~\ref{sec:supp_related_work}.


\noindent\textbf{Density modification and denoising in cryo-EM} Density modification refers to the process of using known properties of the expected density in specific regions of a map to correct errors in observed cryo-EM density maps \citep{terwilliger2020improvement}. Deep learning has increasingly contributed to cryo-EM map denoising and modification, with methods divided into two categories: pretrained models and self-supervised approaches. Pretrained models, like DeepEMhancer \citep{sanchez2021deepemhancer} and EMReady \citep{he2023improvement}, learn the posterior $p(\rvx|\rvy)$ from data to recover high-frequency details. In contrast, self-supervised approaches, such as M \citep{tegunov2021multi} and spIsoNet \citep{liu2024overcoming}, are trained on the dataset being processed, offering more robustness but take longer to process one dataset. 

\noindent\textbf{Missing wedge problem in cryo-ET} Cryo-electron tomography (cryo-ET) is an imaging technique used to reconstruct 3D volumes of biological specimens from 2D images captured at various tilt angles \citep{luvcic2005structural}. The resulting 3D volumes, known as tomograms, provide detailed views of cellular structures, with smaller regions, called subtomograms, extracted to focus on specific structures like proteins \citep{wan2016cryo}. A key challenge in cryo-ET is the missing wedge problem, caused by the limited range of tilt angles during data acquisition, which leaves a wedge-shaped region in Fourier space without information, leading to anisotropic resolution and artifacts. Traditional approaches use signal processing and regularization techniques to address the missing wedge \citep{goris2012electron, deng2016icon, yan2019mbir, zhai2020lottor}, while recent deep learning methods have shown promise in tackling this issue by leveraging data-driven models \citep{liu2022isotropic, van2024missing}. 

\noindent\textbf{\textit{Ab initio} modeling in cryo-EM} \textit{Ab initio} modeling in cryo-EM involves estimating the 3D structure of a protein from 2D particle images with unknown orientations \citep{crowther1970reconstruction}. One of the early approaches involved using 2D class averages, which are representative images created by aligning and averaging particles with similar views to improve the signal-to-noise ratio (SNR) \citep{ludtke1999eman, voss2010toolbox}. However, since the introduction of cryoSPARC, modern methods now use raw particle images directly for model estimation, leveraging stochastic gradient descent (SGD) to bypass the need for class averages \citep{punjani2017cryosparc}. 

\noindent\textbf{Diffusion/Flow Matching for Inverse Problem} Inverse problems refer to a class of problems where the goal is to recover the original sample from \rebuttal{the forward model's} observation, such as image super-resolution \citep{haris2018deep}, inpainting \citep{yeh2017semantic}, deblurring \citep{kupyn2019deblurgan}, etc. 
Since a pretrained DDPM/Flow models the data distribution (prior distribution), it can help discover the posterior distribution given a \rebuttal{forward} model (likelihood). Specifically, \citet{chung2022diffusion} developed diffusion posterior sampling for general inverse problems, removing the need for strong assumptions like the linearity of the \rebuttal{forward} operator. Another line of research examines the solution to the inverse problem when the \rebuttal{forward} operator is unknown \citep{chung2023blind,kapon2024mas}. \rebuttal{For protein structure modeling, \citet{levy2024solving} adopts Chroma \citep{ingraham2023illuminating} as the diffusion prior for a dozen of downstream tasks, including atomic model refinement in cryo-EM.}

\section{Preliminaries}

\rebuttal{In this section, we briefly review two key components of cryoFM. In the training stage, flow matching (\Secref{sec:fm}) is employed as the framework to learn the distribution $p(\mathrm{density})$. In the inference stage, given some observations (noisy densities, 2D projections, etc.), we use the posterior sampling (\Secref{sec:dps}) technique to sample a density from $p(\mathrm{density}|\mathrm{observation})$.}

\subsection{Flow Matching}
\label{sec:fm}
Flow matching \citep{lipman2022flow,liu2022flow} defines the generative process of a data point $\rvx_0 \in \mathbb{R}^n$ by continuously transforming a sample $\rvx_1 \in \mathbb{R}^n$ from the noise distribution, \rebuttal{where the state at an intermediate timestep $t\in[0,1]$ is denoted by $\rvx_t \in \mathbb{R}^n$.}
\rebuttal{A \textit{flow} is the path traversed by the state $\rvx_t$, which} is characterized by an ordinary differential equation (ODE):
\begin{align} \label{eq:flow_ode}
    d\rvx_t = \rebuttal{\nnvtxt}dt,
\end{align}
where \rebuttal{$\nnv:[0,1]\times\mathbb{R}^n\to\mathbb{R}^n$} is a \textit{time-dependent vector field} parameterized by a neural network \rebuttal{parameterized by $\Theta$}, \rebuttal{which defines the transformation of the state $\rvx_t$}.
\rebuttal{At each time step $t$, the  state $\rvx_t$ follows a distribution $p_t(\rvx_t)$, where $p_t:[0,1]\times\mathbb{R}^n\to\mathbb{R}_{\ge 0}$ is a time-dependent probability density function, i.e., $\int p_t(\rvx_t)d\rvx_t=1$. Since $p_t(\rvx_t)$ evolves from the noise distribution $p_1(\rvx_1)$ to the data distribution $p_0(\rvx_0)$, it is referred to as a \textit{probability density path} generated by the vector field \rebuttal{$\nnvtxt$}. An illustration of the concepts of flow matching can be found in \Figref{fig:fm_concepts}.} 

\rebuttal{Given a set of data points sampled from $p_1(\rvx_1)$ and $p_0(\rvx_0)$, flow matching aims to learn the vector field $\nnv$ from the data.} \citet{chen2018neural} directly solves \Eqref{eq:flow_ode} via a differentiable neural ODE solver. \citet{lipman2022flow} and \citet{liu2022flow} suggest that it is more efficient to learn a manually designed conditional vector field $\ru_t(\cdot|\rvx_0):[0,1]\times\mathbb{R}^n\to\mathbb{R}^n$ given a data point $\rvx_0$. For example, the conditional flow can be a straight line between $\rvx_0$ and a noise sample $\rvx_1$:
\begin{align} \label{eq:rec_flow_traj}
    \rvx_t|\rvx_0=(1-t)\rvx_0+t\rvx_1,
\end{align}
where the corresponding vector field is \rebuttal{$\ru_t(\rvx_t|\rvx_0)=\rvx_1-\rvx_0$}.
Finally, the conditional flow matching objective \citep{lipman2022flow} is defined as follows:
\rebuttal{\begin{align*}
  \mathcal{L}(\Theta)=\mathbb{E}_{t \sim \mathcal{U}[0,1],\rvx_0 \sim p_0(\rvx_0),\rvx_1 \sim p_1(\rvx_1)}\|\nnvtxt-(\rvx_1-\rvx_0)\|^2_2,
\end{align*}}
which is equivalent to optimizing the original flow matching objective \citep{lipman2022flow}.

\subsection{Diffusion Posterior Sampling}
\label{sec:dps}
Diffusion posterior sampling (DPS) has emerged as a promising approach for solving inverse problems \citep{song2021scorebased,song2021solving,chung2023solving}. Given a measurement $\rvy\in\mathbb{R}^m$ derived from $\rvx\in\mathbb{R}^n$ with a \rebuttal{forward} operator $\mathcal{A}:\mathbb{R}^n\to\mathbb{R}^m$, the goal is to sample $\rvx$ from the posterior $p(\rvx|\rvy)$. \rebuttal{By Bayes' theorem, it is easy to show that the score of the posterior distribution is:}
\begin{align}\label{eq:bayes-score}
    \nabla_{\rvx_t} \log p_t(\rvx_t|\rvy) = \nabla_{\rvx_t} \log p_t(\rvx_t) + \nabla_{\rvx_t} \log p_t(\rvy|\rvx_t).
\end{align}
However, computing \rebuttal{$\log p_t(\rvy|\rvx_t)=\log \int_{\rvx_0}p(\rvx_0|\rvx_t)p_0(\rvy|\rvx_0)d\rvx_0$} is intractable since it requires the integration over all possible \rebuttal{$\rvx_0 \sim p_0(\rvx_0|\rvx_t)$}. \citet{chung2022diffusion} proposed to use a Laplace approximation of the likelihood term: \rebuttal{$p_t(\rvy|\rvx_{t}) \approx p_0(\rvy|\hat{\rvx}_0(\rvx_t))$}, \rebuttal{where $\hat{\rvx}_0(\rvx_t)$ is the posterior mean estimated by the diffusion model.} The conditional score can thus be approximated by:
\begin{align*}
    \nabla_{\rvx_t} \log p_t(\rvx_t|\rvy) \approx \nabla_{\rvx_t} \log p_t(\rvx_t) + \nabla_{\rvx_t} \log p_0(\rvy|\hat{\rvx}_0(\rvx_t)),
\end{align*}
 If we assume the observation distribution $p_0(\rvy|\rvx)$ is Gaussian conditioned on $\rvx$, \rebuttal{we can show that the gradient of the log-likelihood satisfies}:
\begin{align}\label{eq:Gaussian}
    \nabla_{\rvx_t} \log p_0(\rvy|\hat{\rvx}_0(\rvx_t)) \propto \rebuttal{-}\nabla_{\rvx_t} \|\rvy-\mathcal{A}\hat{\rvx}_0(\rvx_t)\|^2_2.
\end{align}
Putting all these things together, we can approximate the score of the posterior distribution by:
\begin{align*}
    \nabla_{\rvx_t} \log p(\rvx_t|\rvy) \approx \nabla_{\rvx_t} \log p_t(\rvx_t) \rebuttal{-} \lambda_t \cdot \nabla_{\rvx_t} \|\rvy-\mathcal{A}\hat{\rvx}_0(\rvx_t)\|^2_2,
\end{align*}
where $\lambda_t$ is associated with the partition function of Gaussian, and we treat it as a hyperparameter to control the step size of the likelihood term.

\section{\ours}

In this section, we present the implementation of \ours. First, we introduce the pretraining dataset in Section \ref{sec:dataset}. Next, we illustrate the architecture of the neural network in Section \ref{sec:arch}. Finally, we propose a posterior sampling algorithm for the downstream tasks.

\subsection{Pretraining Dataset}
\label{sec:dataset}
Our training dataset consists of deposited sharpened density maps from the EMDB \citep{10.1093/nar/gkad1019}, specifically those with: 1) a reported resolution better than $3.0$~\AA, 2) structures resolved by single-particle cryo-EM, and 3) data entries that include half-maps, ensuring that resolution estimates are based on the ``gold standard" Fourier shell correlation (FSC) \citep{henderson2012outcome}. We manually curated this subset by removing exceptionally large complexes, helical structures, and problematic cases through visual inspection \rebuttal{~(see \Figref{fig:supp_data} for examples)}. We also excluded density maps with side lengths greater than $576$~\AA. This curation resulted in a total of $3479$ density maps, \rebuttal{where $32$ density maps were selected as test set and excluded from training}. The density maps in the selected subset were lowpass filtered to $1.5$~\AA/voxel and $3$~\AA/voxel for \ours-S and \ours-L, respectively. For \ours-S, we applied random cropping to volumes of size $64^3$ along with random rotations for augmentation, whereas data for \ours-L were center cropped to volumes of size $128^3$ and augmented solely with random rotations. \rebuttal{Additionally, we rescale the density value to avoid large numerical variance} \rebuttal{~(\Secref{sec:data_norm} describes the details)}.

\subsection{Architecture}
\label{sec:arch}
\begin{figure}[th!]
\centering
\includegraphics[width=1.0\textwidth]{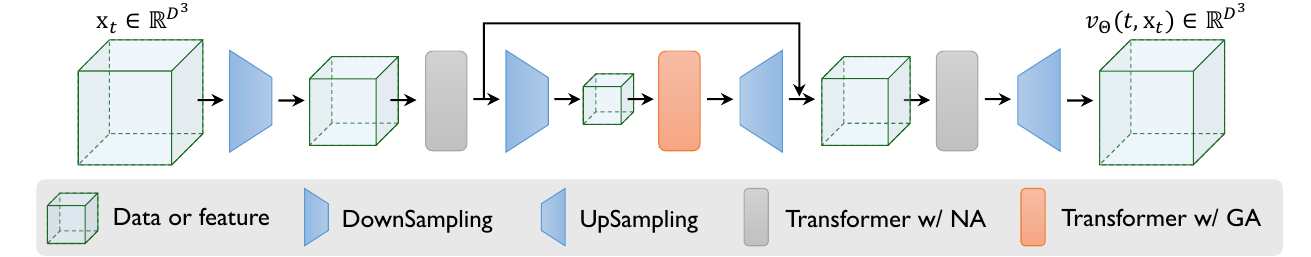}
\caption{\ours's architecture. The side length $D$ of the input $\rvx_t$ undergoes dimension reduction through down-sampling layers and is then expanded back to its original size. To minimize computational cost near the input and output, the model employs neighborhood attention (NA). Neighborhood attention only attend to a localized area, whereas global attention (GA) calculates attention across all positions.}
\label{fig:hdit3d_arch}
\end{figure}

A major challenge in applying the vanilla \rebuttal{Transformer} architecture to 3D density $\rvx_t \in \mathbb{R}^{D^3}$ is the computational complexity, which is $\mathcal{O}(D^6)$ \footnote{The computational complexity of Transformers is $\mathcal{O}(L^2)$, where $L=D^3$ is the number of tokens. }. \ours processes the 3D input data through a Transformer with a \textit{hierarchical} architecture, based on HDiT \citep{crowson2024hdit}. As illustrated in \Figref{fig:hdit3d_arch},  the hierarchical structure downsamples the spatial dimension at initial levels and upsamples it at final levels, implemented via PixelUnshuffle and PixelShuffle layers \citep{shi2016pixelshuffle}. Consequently, the spatial dimension at the middle level is significantly reduced, making the model more easily scalable. Furthermore, at the two ends of the hierarchical structure, \ours employs Neighborhood Attention (NA) \citep{hassani2023natten} since a localized attention layer is able to capture local dependencies while considerably reducing the number of tokens to attend.

\subsection{Flow Posterior Sampling}
\label{sec:fps}


Given a vector field $\nnvtxt$ that generates the prior data distribution $p_0(\rvx_0)$, we aim to convert it to a vector field $\nnvtxty$ that generates the posterior $p_0(\rvx_0|\rvy)$. 
\rebuttal{First, according to \citet{dao2023flow,song2021scorebased}, we can connect the vector field with the score function $\nabla_{\rvx_t} \log p_t(\rvx_t)$ as follows:}
\begin{align*}
    \nnvtxt=f_t(\rvx_t)-\frac{g_t^2}{2}\nabla_{\rvx_t} \log p_t(\rvx_t),
\end{align*}
where $f_t$ is the drift term and $g_t$ is the diffusion coefficient in the forward-time SDE.
\rebuttal{In addition, we define the conditional vector field as follows:}
\begin{align*}
\nnvtxty = f_t(\rvx_t)-\frac{g_t^2}{2}\nabla_{\rvx_t} \log p_t(\rvx_t|\rvy) =\nnvtxt - \frac{g_t^2}{2}\nabla_{\rvx_t}\log p_t(\rvy|\rvx_t),
\end{align*}
\rebuttal{where the second equality holds due to \Eqref{eq:bayes-score}.}
Thus, \rebuttal{substracting the score of the likelihood term} weighted by the diffusion coefficient generates a vector field that models the posterior distribution.
Following the rectified flow \citep{dao2023flow}, we set $f_t(\rvx_t)=-\frac{\rvx_t}{1-t}$ and $\frac{g_t^2}{2}=\frac{t}{1-t}$ in the conditional flow  in \Eqref{eq:rec_flow_traj}, and obtain the conditional vector field as follows:
\begin{align}\label{eq:conditional vector field}
    \nnvtxty = \nnvtxt - \frac{t}{1-t}\nabla_{\rvx_t}\log p_t(\rvy|\rvx_t).
\end{align}

Similar to Section \ref{sec:dps}, by substituting \Eqref{eq:Gaussian} into \Eqref{eq:conditional vector field}, we obtain:
\begin{align*}
    \nnvtxty \approx \nnvtxt + \frac{t}{1-t} \cdot  \lambda_t \nabla_{\rvx_t}  \|\rvy-\mathcal{A}\hat{\rvx}_0(\rvx_t)\|^2_2.
\end{align*}
\rebuttal{where $\hat{\rvx}_0(\rvx_t)=\rvx_t-t\cdot\nnvtxt$ is the posterior mean estimated by the flow model.} In practice, we incorporate some tricks to avoid numerical instability. The detailed algorithm of flow posterior sampling is described in \Algref{alg:fps}. The algorithm can be adapted for different tasks; we refer the reader to Appendix \ref{sec:algorithms} for the additional versions.


\begin{algorithm}[t]
\caption{Flow Posterior Sampling}\label{alg:fps}
\begin{algorithmic}
\Require a pretrained vector field \rebuttal{$\nnv:[0,1]\times\mathbb{R}^n\to\mathbb{R}^n$}, a \rebuttal{forward} operator $\mathcal{A}:\mathbb{R}^n\to\mathbb{R}^m$, an obseravation $\rvy\in\mathbb{R}^m$, number of steps $N$, maximum step size $\lambda_{\text{max}}$ at each step
\Ensure the recovered signal $\rvx_0$
\State $\rebuttal{\Delta t \gets \frac{1}{N}}$
\State $\rvx_1 \gets \mathcal{N}(0,\mI)$
\For {$t \in [1, 1-\rebuttal{\Delta t}, 1-\rebuttal{2 \Delta t}, \cdots, \rebuttal{\Delta t}]$}
    \State $\rvx'_{t-\rebuttal{\Delta t}} \gets \rvx_t- \rebuttal{\Delta t \cdot \nnvtxt}$ 
    \State $l(\rvx_t) \gets \|\rvy-\mathcal{A}\hat{\rvx}_0(\rvx_{t})\|_2^2$
    \State $\rvg \gets \frac{\nabla_{\rvx_{t}} l(\rvx_t)}{\|\nabla_{\rvx_{t} } l(\rvx_t) \|_2}$ \Comment{Normalize the gradient for numerical stability}
    \State $\lambda_t \gets \min\left\{\lambda_{\text{max}}, \frac{t}{1-t}\right\}$ \Comment{Prevent the weighting term from being too large}
    \State $\rvx_{t-\rebuttal{\Delta t}} \gets \rvx'_{t-\rebuttal{\Delta t}} \rebuttal{-} \lambda_t \rvg$ 
\EndFor
\State \Return $\rvx_0$
\end{algorithmic}
\end{algorithm}

\section{Experiments}


Fourier Shell Correlation (FSC) is a widely used metric that compares two density maps in Fourier space \citep{harauz1986exact}, allowing for the assessment of alignment between the ground truth and the reconstructed map. In the experiments, to evaluate the quality of the density maps, we use three primary metrics: $\text{FSC}_{\text{AUC}}$, $\text{FSC}_{0.5}$, and Fail Rate (FR).  Specifically, $\text{FSC}_{\text{AUC}}$ measures the overall correlation across all spatial frequencies, and $\text{FSC}_{0.5}$ represents the resolution of the reconstructed map at the standard 0.5 cutoff \citep{rosenthal2003optimal}. The Fail Rate (FR) identifies cases where the method fails to run or produces a result that significantly deviates from the ground truth. For the FSC metrics, we only report the results for cases that did not fail. More detailed explanation can be found in Section \ref{sec:supp_metrics}.



\subsection{Spectral Noise Denoising}\label{sec:spectral_noise}

In cryo-EM reconstruction, spectral noise is the most commonly used noise model due to its ability to capture the varying noise characteristics across different spatial frequencies, which arise from factors such as the contrast transfer function (CTF) and detector imperfections.
We consider the introduction of noise in the Fourier domain, where the variance of the added Gaussian noise is frequency-dependent, with higher frequencies exhibiting greater noise variance. Given a density $\fV \in \mathbb{C}^{D\times D\times D}$ in the Fourier domain, the \rebuttal{forward} model $\mathcal{A}: \mathbb{C}^{D\times D\times D} \to \mathbb{C}^{D\times D\times D}$ is:
\begin{align*}
    \mathcal{A}(\fV) = \fV + \rvepsilon,
\end{align*}
where $\rvepsilon \in \mathbb{R}^{D\times D\times D}$ is a noise volume, whose values on the spherical shell with the same radius $\nu$ are the same ($\nu$ denotes the index of a component in the frequency space):
\begin{align}
\label{eq:noise_power}
    \quad \rvepsilon(\nu) \sim \mathcal{N}(0, \sigma_{\text{noise}}^2(\nu)).
\end{align}
In the experiment, we estimate $\sigma_{\text{noise}}^2(\nu)$ from the half maps, which are two independent observations with the same underlying signal with uncorrelated noise. See Section \ref{sec:spectral_noise_estimation} for details.



\begin{table}[h]
\caption{Results of the spectral noise denoising task, comparing \ours with DeepEMhancer \citep{sanchez2021deepemhancer} and EMReady \citep{he2023improvement}. \rebuttal{In Sections \ref{sec:spectral_noise}, \ref{sec:anisotropic_noise}, \ref{sec:missing_wedge}, the forward model acts as a degradation of the original data.} The best attainable $\text{FSC}_{0.5}$ is $3.0$ \AA~since the voxel size is $1.5$ \AA.}\label{tab:spectral_noise_results}
\centering
\resizebox{\textwidth}{!}{%
\begin{tabular}{lccccccccccccc}
\toprule
{} & \multicolumn{9}{c}{Estimated resolution at different level of noise} \\
\cmidrule{2-10}
& \multicolumn{3}{c}{$3.2$~\AA} & \multicolumn{3}{c}{$4.3$~\AA} & \multicolumn{3}{c}{$6.1$~\AA} \\
\cmidrule(r){2-4} \cmidrule(r){5-7} \cmidrule(r){8-10}
{} & FR$\downarrow$ & $\text{FSC}_{\text{AUC}}\uparrow$ & $\text{FSC}_{0.5}\downarrow$ & FR$\downarrow$ & $\text{FSC}_{\text{AUC}}\uparrow$ & $\text{FSC}_{0.5}\downarrow$ & FR$\downarrow$ & $\text{FSC}_{\text{AUC}}\uparrow$ & $\text{FSC}_{0.5}\downarrow$ \\
\cmidrule(r){1-1} \cmidrule(r){2-4} \cmidrule(r){5-7} \cmidrule(r){8-10}

After degradation & - & $0.8876$ & $3.21$ & - & $0.5527$ & $4.96$ & - & $0.4969$ & $6.07$ \\
\cmidrule(r){1-1} \cmidrule(r){2-4} \cmidrule(r){5-7} \cmidrule(r){8-10}

DeepEMhancer & $0.06$ & $0.92\pm0.02$ & $\mathbf{3.0\pm0.0}$ & $0.34$ & $0.66\pm0.05$ & $\mathbf{4.3\pm0.2}$ & $0.44$ & $\mathbf{0.62\pm0.05}$ & $\mathbf{4.5\pm0.3}$ \\

EMReady & $0.03$ & $0.73\pm0.02$ & $3.4\pm0.1$ & $0.16$ & $0.59\pm0.08$ & $4.9\pm0.9$ & $0.19$ & $0.55\pm0.09$ & $5.3\pm0.9$ \\

\ours & $\mathbf{0.00}$ & $\mathbf{0.95\pm0.01}$ & $\mathbf{3.0\pm0.0}$ & $\mathbf{0.00}$ & $\mathbf{0.68\pm0.04}$ & $4.4\pm0.2$ & $\mathbf{0.00}$ & $\mathbf{0.62\pm0.04}$ & $4.6\pm0.3$ \\
\bottomrule
\end{tabular}%
}
\end{table}

\begin{figure}[h]
\centering
\includegraphics[width=0.9\textwidth]{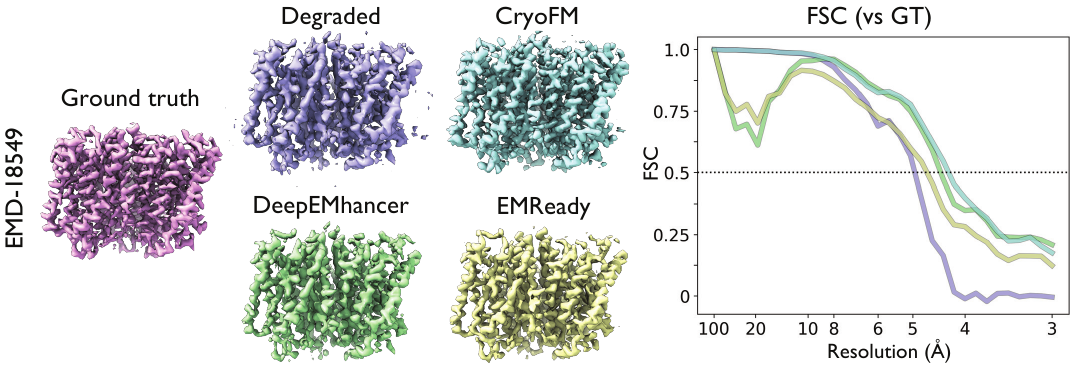}
\caption{Result of the spectral noise denoising task. Two density maps from EMDB were added spectral noise so that the estimated resolution is $4.3$~\AA. The degraded density maps \rebuttal{(results after applying the forward model)} were filtered by \textit{relion\_postprocess} for visual clarity.}
\label{fig:spectral_noise_result}
\end{figure}

We manually introduce spectral noise, resulting in resolutions ranging from 3.2~\AA{} to 15.0~\AA{}. DeepEMhancer and EMReady, both pretrained models that directly learn the posterior $ p(\rvx|\rvy) $, are used as baselines. \ours demonstrates superior robustness, with no failed cases across all experiments, while both baselines show higher fail rates, especially at medium resolutions. Additionally, as shown in \Tabref{tab:spectral_noise_results} and \Tabref{tab:supp_spectral_noise_results} in Section \ref{sec:spectral_noise_estimation}, \ours achieves higher $\text{FSC}_{\text{AUC}}$ and better $\text{FSC}_{0.5}$ in most cases, significantly improving degraded density maps and enhancing the resolution of helices and loops, as illustrated in \Figref{fig:spectral_noise_result}. Notably, the FSC curves demonstrate that \ours successfully improves the signal across all frequencies, while the baselines fails to retain trustworthy signals in the low-frequency range.


\subsection{Anisotropic Noise Denoising}\label{sec:anisotropic_noise}





\begin{wrapfigure}{r}{0.45\textwidth} 
    \centering
    \includegraphics[width=\linewidth]{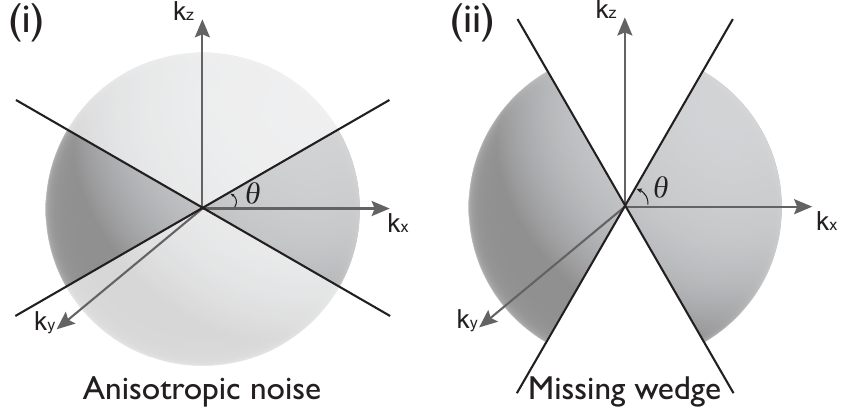}
    \caption{\rebuttal{Forward} operators for (i) the anisotropic noise and (ii) the missing wedge.}
    \label{fig:degrade_op}
\end{wrapfigure}

Anisotropic noise in cryo-EM occurs when noise distribution varies by direction, affecting some orientations more than others and leading to uneven reconstruction quality. As a result, certain orientations have lower signal-to-noise ratios than others. To approximate this degradation in a simplified manner, we amplify the spectral noise by a factor when the particle orientation falls within a specific range of angles, simulating the increased uncertainty in less-sampled orientations. The \rebuttal{forward} model $\mathcal{A}: \mathbb{C}^{D\times D\times D} \to \mathbb{C}^{D\times D\times D}$ is given by:
\begin{align*}
    \mathcal{A}(\fV)= 
    \begin{cases}
    \fV + \rvepsilon, & \text{if } \theta_{\min} \leq \theta \leq \theta_{\max} \\
    \fV + \alpha\cdot\rvepsilon, & \text{otherwise}
    \end{cases}, 
\end{align*}
where $\rvepsilon \in \mathbb{R}^{D\times D\times D}$ is the same as defined in \Eqref{eq:noise_power}, and $\alpha > 1$ is a \rebuttal{scalar} that increases the noise for specific orientation angles, reflecting the heightened uncertainty in those directions. $\theta_{\min}$ and $\theta_{\max}$ are the angles that controls the portion of the noise being amplified as illustrated in \Figref{fig:degrade_op}. In the experiment, we estimate $\sigma_{\text{noise}}^2(\nu)$ and $\alpha$ from the half maps. See Section \ref{sec:aniso_noise_estimation} for details.



\begin{table}[t]
\caption{Results of the anisotropic noise denoising task, comparing \ours with DeepEMhancer \citep{sanchez2021deepemhancer} and spIsoNet \citep{liu2024overcoming}.}\label{tab:anisotropic_noise_results}
\centering
\resizebox{\textwidth}{!}{%
\begin{tabular}{lccccccccc}
\toprule
\cmidrule{2-10}
& \multicolumn{3}{c}{$\theta \in [-45^\circ, +45^\circ]$} & \multicolumn{3}{c}{$\theta \in [-30^\circ, +30^\circ]$} & \multicolumn{3}{c}{$\theta \in [-15^\circ, +15^\circ]$} \\

\cmidrule(r){2-4} \cmidrule(r){5-7} \cmidrule(r){8-10}
{} & FR$\downarrow$ & $\text{FSC}_{\text{AUC}}\uparrow$ & $\text{FSC}_{0.5}\downarrow$ & FR$\downarrow$ & $\text{FSC}_{\text{AUC}}\uparrow$ & $\text{FSC}_{0.5}\downarrow$ & FR$\downarrow$ & $\text{FSC}_{\text{AUC}}\uparrow$ & $\text{FSC}_{0.5}\downarrow$ \\ 
\cmidrule(r){1-1} \cmidrule(r){2-4} \cmidrule(r){5-7} \cmidrule(r){8-10}

After degradation & $-$ & $0.6623$ & $4.15$ & $-$ & $0.6324$ & $4.27$ & $-$ & $0.6111$ & $4.38$ \\
\cmidrule(r){1-1} \cmidrule(r){2-4} \cmidrule(r){5-7} \cmidrule(r){8-10}

DeepEMhancer & $0.22$ & $0.80\pm0.03$ & $3.2\pm0.1$ & $0.22$ & $0.79\pm0.05$ & $3.3\pm0.2$ & $0.22$ & $0.77\pm0.05$ & $\mathbf{3.4\pm0.2}$ \\

spIsoNet & $\mathbf{0.00}$ & $0.65\pm0.01$ & $4.15\pm0.01$ & $\mathbf{0.00}$ & $0.62\pm0.01$ & $4.27\pm0.03$ & $\mathbf{0.00}$ & $0.61\pm0.01$ & $4.37\pm0.03$ \\

\ours & $\mathbf{0.00}$ & $\mathbf{0.88\pm0.03}$ & $\mathbf{3.1\pm0.1}$ & $\mathbf{0.00}$ & $\mathbf{0.84\pm0.03}$ & $\mathbf{3.2\pm0.1}$ & $\mathbf{0.00}$ & $\mathbf{0.81\pm0.04}$ & $\mathbf{3.4\pm0.2}$ \\
\bottomrule
\end{tabular}%
}
\end{table}

\begin{figure}[h]
\centering
\includegraphics[width=0.9\textwidth]{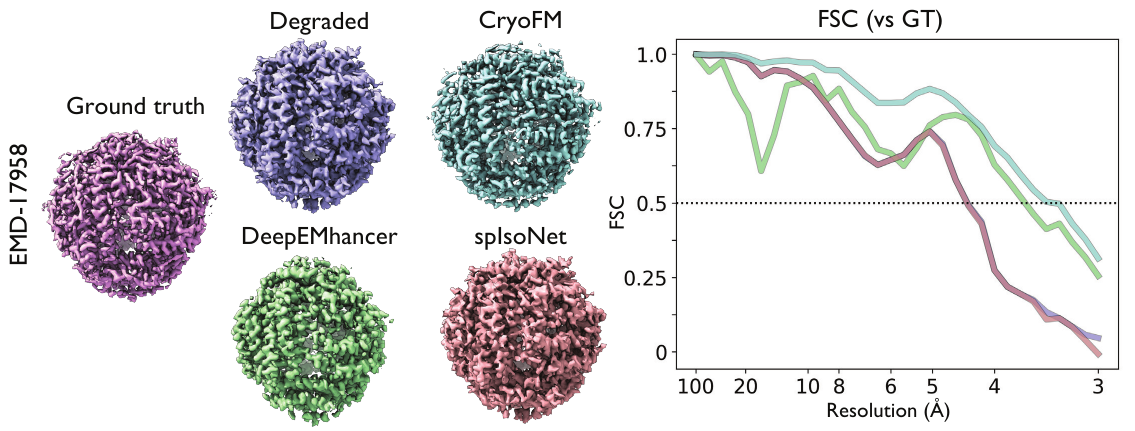}
\caption{Result of the anisotropic noise denoising task. Two density maps from EMDB are added anisotropic spectral noise that the estimated resolution is 4.38 \AA. The degraded density maps are filtered by \textit{relion\_postprocess} for visual clarity.}
\label{fig:anisotropic_noise_result}
\end{figure}

We manually add varied anisotropic spectral noise to the test set density maps and use DeepEMhancer, spIsoNet, and \ours for signal restoration. As shown in \Tabref{tab:anisotropic_noise_results}, \ours outperforms both baselines on FSC metrics and has a zero fail rate. \Figref{fig:anisotropic_noise_result} demonstrates that \ours restores the corrupted signal without introducing significant bias, unlike DeepEMhancer, which introduces artifacts and alters the overall shape. While spIsoNet preserves signals across frequencies, it fails to improve correlation with the ground truth. This suggests that pretrained models that learn the posterior $ p(\rvx|\rvy) $ like DeepEMhancer can struggle to retain reliable signals, and self-supervised methods like spIsoNet lack a data-driven foundation, limiting their effectiveness. Additionally, spIsoNet requires over 5 hours on a V100 GPU for one dataset, whereas \ours completes the task in about 30 minutes.


\subsection{Missing Wedge Restoration}\label{sec:missing_wedge}

The missing wedge problem in cryo-ET arises from the limited tilt range of the electron microscope during data acquisition, causing incomplete Fourier space sampling and artifacts in the subtomograms. We simulate this effect by applying a wedge-shaped mask in the Fourier domain, removing data from unmeasured orientations. The \rebuttal{forward} model $\mathcal{A}: \mathbb{C}^{D\times D\times D} \to \mathbb{C}^{D\times D\times D}$ is given by:
\begin{align*}
    \mathcal{A}(\fV)= 
    \begin{cases}
    \tilde{\boldsymbol{V}}, & \text{if } \theta_{\min} \leq \theta \leq \theta_{\max} \\
    0, & \text{otherwise}
    \end{cases},
\end{align*}
where $\theta_{\min}$ and $\theta_{\max}$ are typically set to $-60^\circ$ and $+60^\circ$ respectively, representing the common tilt angle limits in experimental setups as illustrated in \Figref{fig:degrade_op}. For posterior sampling, we slightly modify the flow posterior sampling algorithm to get \Algref{alg:missing_wedge}. \rebuttal{The modification is based on the fact that it is not necessary to compute the gradient of the loss function, since we can directly combine the observed part in $\rvy$ with the remaining part in $\rvx_t$ to maximize the likelihood.}

\begin{wraptable}{r}{0.55\textwidth}
\caption{Results of the missing wedge restoration task.}\label{tab:missing_wedge_result}
\centering
\begin{tabular}{lcc}
\toprule
{} & $\text{FSC}_{\text{AUC}}\uparrow$ & \makecell[c]{$\text{FSC}_{\text{AUC}}\uparrow$ \\ {\scriptsize (Missing Region)}} \\
\cmidrule(r){1-1} \cmidrule(r){2-3}
After degradation & $0.80\pm0.02$ & $0.0000$ \\
\cmidrule(r){1-1} \cmidrule(r){2-3}
\ours & $0.92\pm0.02$ & $0.76\pm0.06$ \\
\bottomrule
\end{tabular}
\end{wraptable}

\begin{figure}[t]
\centering
\includegraphics[width=0.9\textwidth]{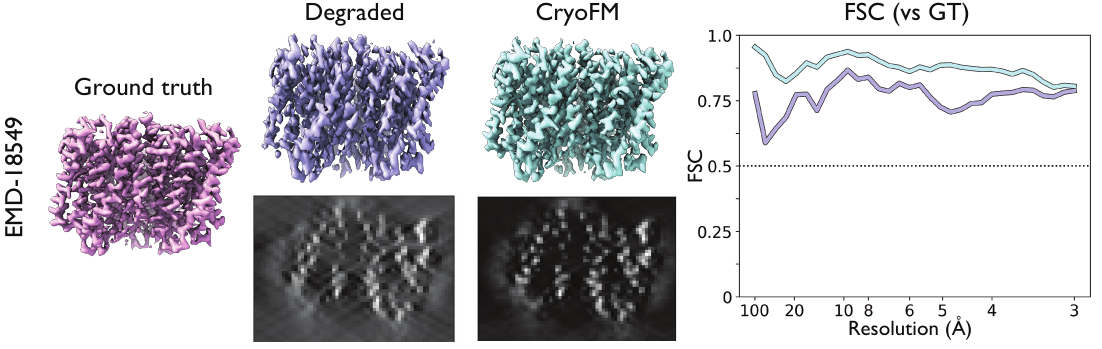}
\caption{Result of the missing wedge restoration task. Two maps from EMDB are masked in Fourier space to simulate the missing wedge effect in cryo-ET. The central slices through the maps are also shown.}
\label{fig:missing_wedge_result}
\end{figure}

\begin{figure}[ht]
\centering
\includegraphics[width=0.9\textwidth]{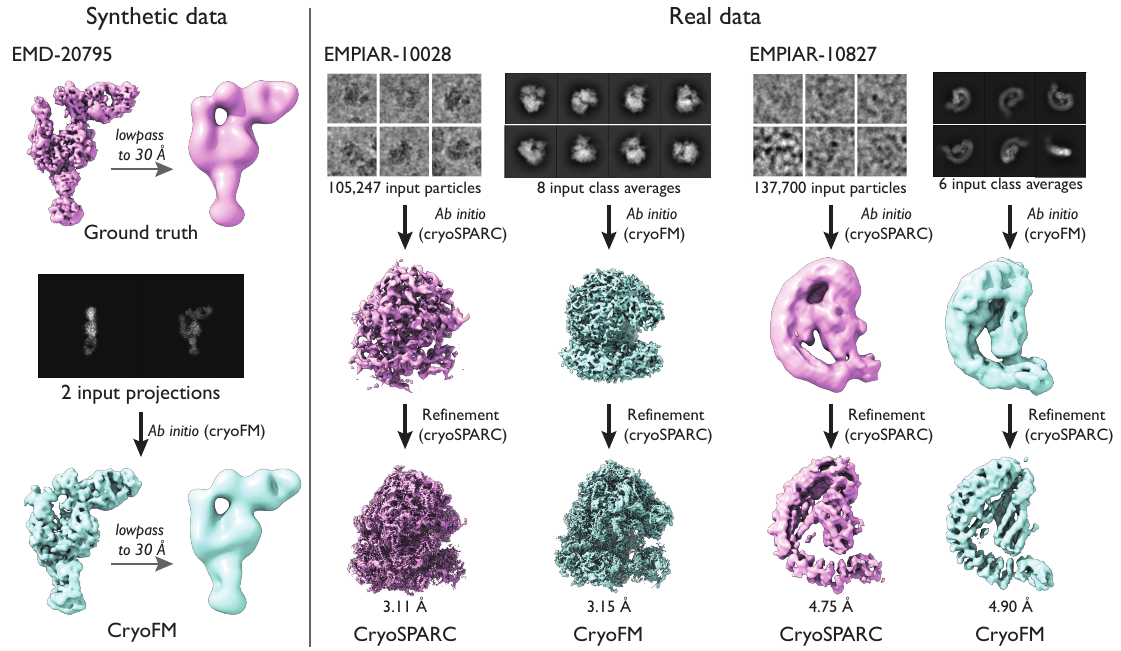}
\caption{Result of the \textit{ab initio} modeling task. One synthetic data and two real data are shown.}
\label{fig:abinitio_result}
\end{figure}

We apply the missing wedge effect to 32 test set density maps, with results in \Tabref{tab:missing_wedge_result} and an example in \Figref{fig:missing_wedge_result}. \ours effectively restores the original signal, reducing the missing wedge artifacts, as shown by both FSC metrics and visual inspection. The FSC curve confirm a strong correlation with the ground truth, particularly in the low-frequency range.


\subsection{\textit{Ab initio} Modeling}


\textit{Ab initio} modeling in cryo-EM involves reconstructing a coarse density map from 2D particle projections, which serves as a reference for refinement. Here, we simplify the problem by focusing on generating a coarse density map from a few clean 2D projections obtained via upstream 2D classification. This approach mirrors earlier \textit{ab initio} methods \citep{ludtke1999eman, voss2010toolbox} before cryoSPARC’s introduction of SGD \citep{punjani2017cryosparc}. Given $K$ projections, there exist $K$ \rebuttal{forward} operators, where the $k$-th \rebuttal{forward} model is $\mathcal{A}^{(k)}: \mathbb{R}^{D\times D\times D} \to \mathbb{R}^{D\times D}$ given by:
\begin{align*}
    \mathcal{A}^{(k)}(\mV) = \mathcal{P}(\phi^{(k)}, \mV),~~~ k\in[1, 2, \cdots, K],
\end{align*}
where $\mathcal{P}$ is a projection operator in the real space, and $\phi^{(k)}\in\mathbb{SO}(3)\times\mathbb{R}^2$ is the pose of the 2D projections. In this task, $\phi^{(k)}$ is unknown and can not be pre-determined. \rebuttal{We modify the flow posterior sampling to get \Algref{alg:ab_initio}, where the main modifications are: (i) searching the optimal pose iteratively in the sampling process \citep{scheres2012relion1,punjani2017cryosparc,zhong2021cryodrgn2}, and (ii) computing the likelihood using correlation to avoid numerical issues.} Moreover, since \textit{ab initio} modeling focuses on capturing the global shape at low resolution, we use \ours-L for this task.

We apply \ours to one synthetic dataset and two real datasets from the Electron Microscopy Public Image Archive (EMPIAR). For the synthetic dataset, we use 2 clean projections, while for the real datasets, we use 8 or 6 selected class averages as the input. As shown in \Figref{fig:abinitio_result}, \ours closely matches the ground truth at low frequencies on the synthetic dataset. For the real datasets, \ours produces \textit{ab initio} models that achieve similar final resolutions after refinement as those generated by cryoSPARC, demonstrating its potential for 3D reconstruction in cryo-EM.

\subsection{Ablation \& Discussion}


In this section, we perform ablation studies to analyze the impact of different design choices and hyper-parameters. We briefly present some conclusions here and refer the reader to Appendix \ref{sec:supp_ablate}.

\begin{figure}[ht]
    \centering
    \begin{minipage}{0.56\linewidth} 
        \centering
        \includegraphics[width=\textwidth]{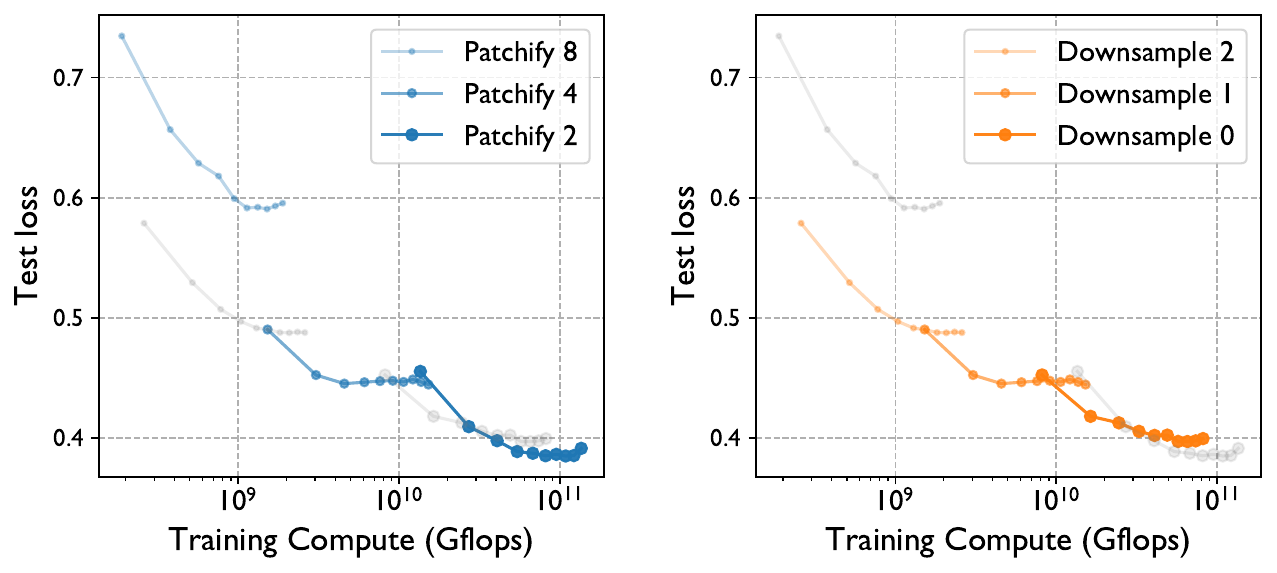}
        \caption{MSE loss on test set as a function of total training compute. The estimation of training compute is consistent with \citet{peebles2023dit}.}
        \label{fig:ablate_model_arch}
    \end{minipage}
    \hspace{0.03\linewidth}
    \begin{minipage}{0.26\linewidth} 
        \centering
        \includegraphics[width=\textwidth]{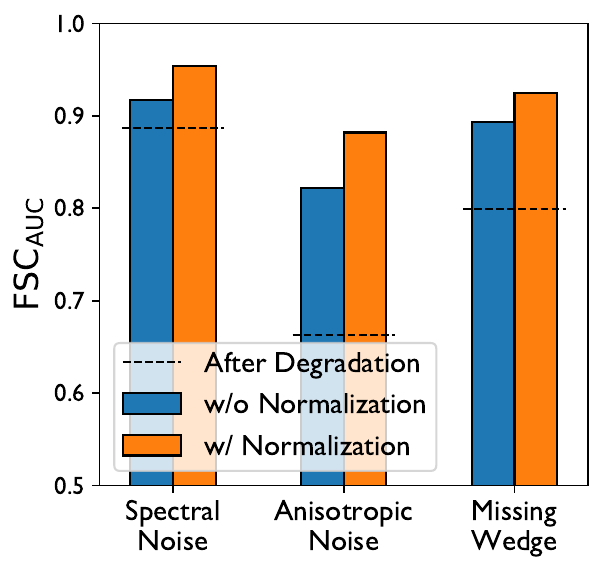}
        \caption{Downstream performance with respect to data normalization.}
        \label{fig:ablate_norm}
    \end{minipage}
\end{figure}

\begin{figure}[th]
    \centering
    \begin{minipage}[b]{0.27\linewidth} 
        \centering
        \includegraphics[width=\textwidth]{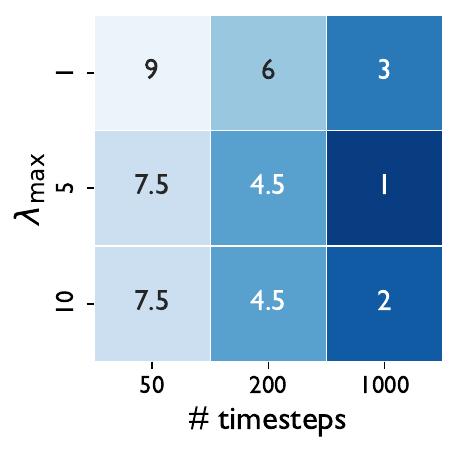}
        \caption{Averaged ranking ($\downarrow$) across downstream tasks for hyperparameters.}
        \label{fig:ablate_dps_hyp}
    \end{minipage}
    \hspace{0.05\linewidth}
    \begin{minipage}[b]{0.54\linewidth} 
        \centering
        \includegraphics[width=1.0\textwidth, trim=0 -12pt 0 0, clip]{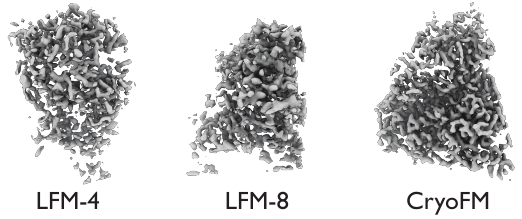}
        \caption{Unconditional sampling from models in different spaces. For latent flow models (LFM), the high-frequency information contains more noise.}
        \label{fig:ldm_results}
    \end{minipage}
\end{figure}

\noindent\textbf{Moderate patchifying and downsampling parameters lead to efficient training with minimal performance degradation.} As shown in \Figref{fig:ablate_model_arch}, reducing both patchifying and downsampling parameters lowers the test loss, since less downsampling causes less information loss. However, the reduction becomes marginal while the training compute (Gflops) increases significantly. To achieve a balance between the performance and training cost, we set the patchifying parameter to $4$ and the downsampling parameter to $1$, respectively.


\noindent\textbf{Data normalization enhances downstream performance across all tasks.} \rebuttal{Data normalization is the last step in data processing (as referenced in \Secref{sec:data_norm}). A density with a value ranging from $[0, 1]$ is subtracted by $0.04$ and then divided by $0.09$.} \Figref{fig:ablate_norm} demonstrates that models trained with normalized data consistently improves the $\text{FSC}_{\text{AUC}}$ metrics compared to the one without normalization in three downstream tasks. The observation is in line with some DDPM-based work \citep{yim2023se,rombach2022high}, as a significant change in the variance of the data can make the model difficult to train \citep{karras2022elucidating}.

\noindent\textbf{More sampling steps boosts the performance in density restoration.} \Figref{fig:ablate_dps_hyp} presents the averaged ranking across three downstream tasks for different time steps ($\#\mathrm{timesteps}$) and the maximum step sizes ($\lambda_\text{max}$). The main factor influencing performance is the number of time steps. A smaller $\lambda_\text{max}$ may reduce the impact of the likelihood term and thus degrade performance. We choose the best combination: $\lambda_\text{max} = 5$ and $\#\mathrm{timesteps}=1000$.


\noindent\textbf{Flow models trained in the voxel space converge much faster and better than the latent flow models.}
We conducted experiments on training the model in the latent space \citep{rombach2022high}. \Figref{fig:ldm_results} illustrates the unconditional sampling results from two latent flow models (LFM), where the high-frequency information is more noisy. The observation concides with that of \citet{crowson2024hdit}, where they found that latent diffusion models fail to generate fine details. We suspect that the VAE is under-trained due to a small amount of data, \rebuttal{and we will leave this for future work}.

\section{Conclusion}

In this study, we present \ours, a flow matching-based foundation model that learns the \textit{prior} distribution of high-quality cryo-EM densities. During inference, we derive the \rebuttal{forward} operators for specific tasks, allowing protein densities to be sampled from the posterior distribution based on given observations. \ours demonstrates versatility by restoring protein densities across four distinct tasks without fine-tuning, showcasing the potential of deep generative models as a powerful prior in cryo-EM. While promising, \ours has limitations: though we use real data with synthetic noise for comparison, applying the method directly to real-world noisy densities remains challenging. Additionally, our work does not address reconstructing 3D densities from raw 2D particles, but we believe \ours can contribute to solving these complex tasks in future work.

\bibliography{reference}
\bibliographystyle{iclr2025_conference}

\appendix


\rebuttal{\section{Discussion on Flow Matching}}

\rebuttal{\subsection{Concepts in Flow Matching}}

\begin{figure}[ht]
\centering
\includegraphics[width=0.8\textwidth]{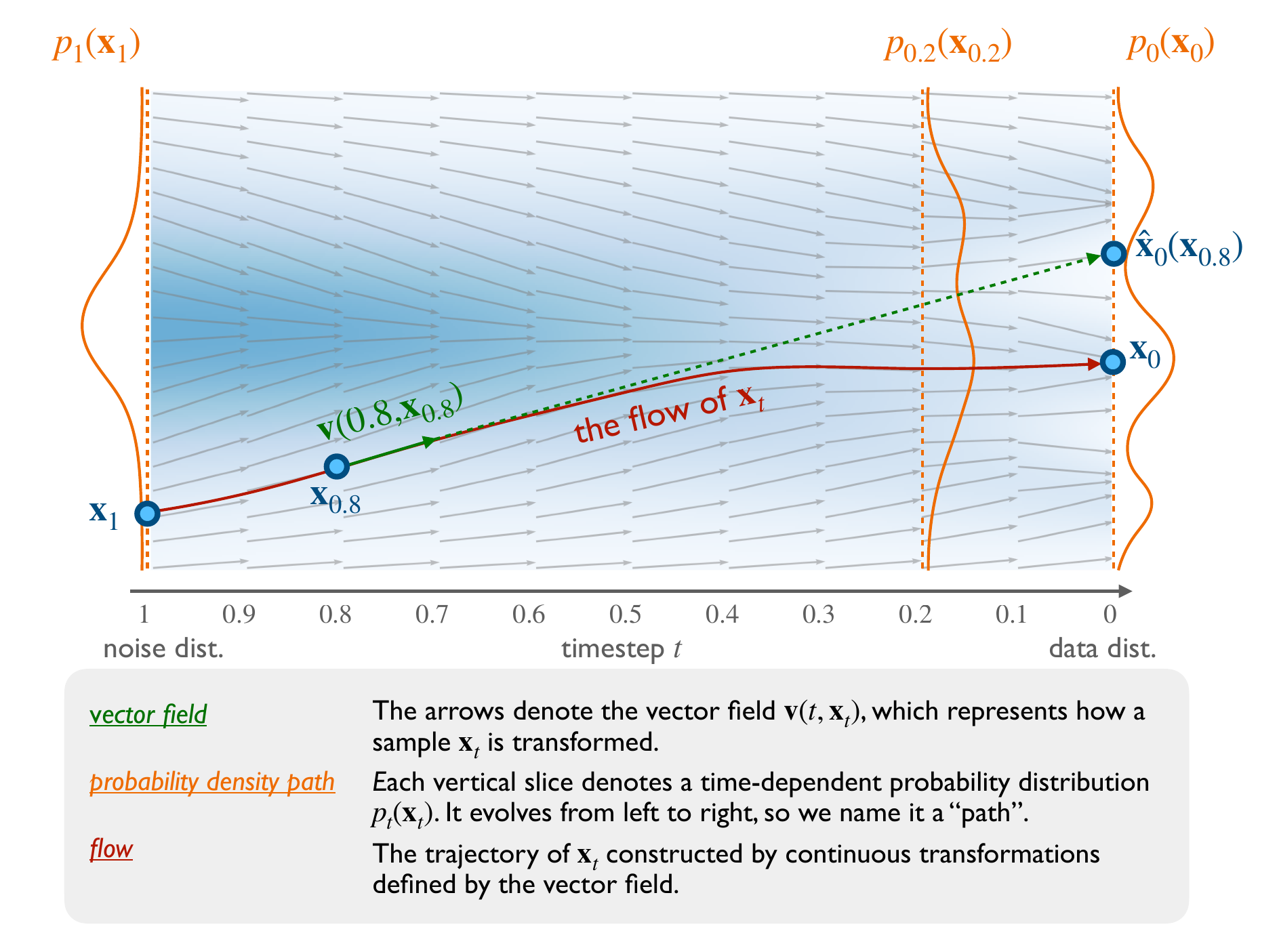}
\caption{\rebuttal{An illustration of three main concepts in flow matching: (i) the vector field $\nnvtxt$, (ii) the probability density path $p_t(\rvx_t)$, and (iii) the flow of $\rvx_t$. The approximated posterior mean $\hat{\rvx}_0(\rvx_t)$ in \Algref{alg:fps} is also depicted on the right part.}}
\label{fig:fm_concepts}
\end{figure}

\rebuttal{\Figref{fig:fm_concepts} illustrates three main concepts in flow matching. Here, the horizontal bar represents the time dimension, and each vertical slice represents a time-dependent probability distribution with only one dimension.}

\rebuttal{\subsection{Choice of Flow Matching for Foundation Models}}
\rebuttal{In this section, we discuss the choice of employing flow matching as the backbone of the cryo-EM foundation model. The main motivation is that (i) generative modeling is a trend in building modern foundation models, and (ii) flow matching excels at generative modeling.}

\rebuttal{\textbf{Generative modeling is a trend in building modern foundation models.} ChatGPT \citep{achiam2023gpt} has disrupted the entire NLP (natural language processing) area with its extremely strong generative power. It brings about a paradigm shift from \textit{representation learning} to \textit{generative modeling}. For representation learning, one of the most prominent models is BERT \citep{devlin2018bert}, which performs well in language understanding but is not adept at performing generation. Similar trends can also be observed in computational biology. Two typical cases in this field are ESM-3 and AlphaFold3. Built upon ESM \citep{rives2021biological} (a BERT-style model), \citet{hayes2024simulating} developed ESM-3 that has the ability to generate protein sequences. A major change from AlphaFold2 \citep{jumper2021highly} to AlphaFold3 \citep{abramson2024accurate} is the diffusion module that can be utilized to generate an ensemble of structures.}

\rebuttal{\textbf{Flow matching excels at generative modeling.} Diffusion-based foundation models have had a significant impact on generative modeling in AI for biology. AlphaFold3 \citep{abramson2024accurate}, RFDiffusion \citep{watson2023novo}, and Chroma \citep{ingraham2023illuminating} are all based on diffusion models (DDPM). ESM-3 \citep{hayes2024simulating}, although not explicitly stated in its manuscript, is closely related to discrete diffusion models. One of the crucial reasons for the success of DDPM resides in iterative refinement, which gradually refines the generated samples to a more superior state. The iterative nature of DDPM enables it to outperform its counterparts with one-step generation methods, such as VAE and GAN. Flow matching can be regarded as a variant of DDPM and shares many similar properties with DDPM. We prefer flow matching over DDPM since it is easier to train and can converge faster. In the early stage of this study, we trained both DDPM and flow matching. We found that flow matching can yield similar results while incurring less computational cost.}

\section{Additioiant onal Related Work}
\label{sec:supp_related_work}

\subsection{Density modification and denoising in cryo-EM}
\label{sec:supp_related_work_densitymod}

Traditional density modification methods often apply Fourier space weighting, utilizing frequency-dependent scaling to suppress noise and enhance signal \citep{jakobi2017model, ramirez2020automatic, terwilliger2020cryo, kaur2021local}. These heuristic-based approaches, like Wiener-style deblurring \citep{ramirez2020automatic}, tend to rely on local resolution estimates and filtering strategies but may struggle with intricate structural details due to limited priors. On the deep learning side, Blush \citep{kimanius2024data} further refine density maps by denoising half-maps during iterative refinement. However, they can introduce hallucinated details in lower-resolution maps. In contrast, methods such as M \citep{tegunov2021multi} and spIsoNet \citep{liu2024overcoming} avoid external data, with M leveraging noise2noise \citep{moran2020noisier2noise} for denoising independent half-maps, and spIsoNet addressing anisotropic signal distributions through self-supervised learning. Though slower, these approaches tend to reduce the risk of hallucinations and offer better robustness, but their power is often limited by not being data-driven. In this paper, we selected DeepEMhancer \citep{sanchez2021deepemhancer}, EMReady \citep{he2023improvement} and spIsoNet as the baselines for the denoising tasks.

\rebuttal{\textbf{DeepEMhancer} DeepEMhancer is a deep learning-based tool designed to enhance cryo-EM density maps, improving their interpretability and aiding in structural analysis. The method employs a convolutional neural network (CNN) trained on pairs of experimental cryo-EM maps and corresponding locally sharpened maps. This training enables DeepEMhancer to learn the mapping between low-quality input maps and their high-quality counterparts. The pretrained model processes new cryo-EM maps to produce enhanced versions with improved clarity and detail, facilitating more accurate structural interpretations.}

\rebuttal{\textbf{EMReady} Similar to DeepEMhancer, EMReady is a computational method designed to enhance cryo-EM density maps by simultaneously applying local and non-local denoising techniques. The method uses a Swin-Conv-3DUNet and trained on pairs of experimental cryo-EM maps and correponding synthetic maps from the atomic models. The pretrained model can enhance cryo-EM maps for better interpretations.}

\rebuttal{\textbf{spIsoNet} spIsoNet is a self-supervised deep learning method developed to address the preferred orientation problem in single-particle cryo-EM. This issue arises when particles predominantly adopt specific orientations during imaging, leading to anisotropic data and potential inaccuracies in 3D reconstructions. The method adopts a U-Net architecture and leverages the half maps in cryo-EM reconstruction to learn representations from the overrepresented orientations, recovering information for the underrepresented or missing views.}

\subsection{Missing wedge problem in cryo-ET}
\label{sec:supp_related_work_misswedge}

Cryo-ET enables detailed visualization of macromolecular complexes and cellular structures in their native environments. The missing wedge problem occurs due to the physical limitation of tilt angles during data acquisition, typically restricted to $(-60^\circ, +60^\circ)$, resulting in missing data in Fourier space and artifacts such as elongation along the missing axes. Traditional methods that use signal processing techniques and regularization strategies \citep{goris2012electron, deng2016icon, yan2019mbir, zhai2020lottor} are often based on heuristic assumptions and have limitations in fully recovering lost data. Recently, deep learning models have been applied to this problem, offering improved recovery of complex patterns in tomograms \citep{liu2022isotropic, van2024missing}. However, these methods often focus on entire tomograms, making it difficult to incorporate specific prior knowledge of protein structures, limiting their effectiveness in subtomogram reconstructions.

\subsection{\textit{Ab initio} modeling in cryo-EM}
\label{sec:supp_related_work_abinitio}

Earlier approaches to \textit{ab initio} modeling relied on experimental techniques, such as image tilt pairs \citep{radermacher1986new, leschziner2006orthogonal}, which provided indirect pose information, or negative stain \citep{de2011negative}, which improved SNR but at the cost of high-frequency detail. Computationally, 2D class averages were commonly used as input due to their higher SNR compared to raw particles, though they had limitations, particularly in fully sampling Fourier space due to the restricted range of particle orientations \citep{voss2010toolbox}. While cryoSPARC's SGD approach improved this by working directly with raw particles \citep{punjani2017cryosparc}, certain structural features or heterogeneity observed in 2D class averages may still be lost in the final 3D reconstruction, especially in challenging samples.

\subsection{Diffusion/Flow-based Model}
\label{sec:supp_related_work_diffusion}

Flow-based models \citep{lipman2022flow,liu2022flow} and denoising diffusion probablistic models (DDPM) \citep{sohl2015deep,ho2020denoising} are two modern deep generative models. They exhibit many similarities in technical details. DDPM implements an iterative refinement process by learning to gradually denoise a sample from a normal distribution. It has achieved the state-of-the-art results on many generative tasks, including image generation \citep{rombach2022high,podell2023sdxl,peebles2023dit}, video generation \citep{blattmann2023stable}, and molecule generation \citep{yim2023se,abramson2024accurate,ingraham2023illuminating,watson2023novo,wang2024diffusion}, etc. Recently, diffusion models have been adopted in the cryo-EM field. \citet{kreis2022latent} traverses the latent space of cryoDRGN \citep{zhong2021cryodrgn} with a diffusion model, while \rebuttal{\citet{wang2024diffmodeler}} refines structures for model building by iteratively denoising the density. Flow-based models regress a vector field that generate a disired probability path. Their simple and efficient implementation enables fast learning. These models have shown success in various domains, including image generation \citep{esser2024scaling} and moleculer generation \citep{yim2023se,bose2023se}. The DDPM objective can also be unified into flow-based models by converting it into a probability flow ODE \citep{song2021scorebased}.

\subsection{Vision Transformers for Diffusion Models}
\label{sec:supp_related_work_transformer}
Diffusion transformers \citep{peebles2023dit} have demonstrated significant scalability and generative capabilities in image-related tasks \citep{esser2024scaling,hoogeboom2023simple,hatamizadeh2023diffit,zhou2024transfusion}. In particular, HDiT \citep{crowson2024hdit} leverages the inherent hierarchical nature of visual patterns in its model design. By integrating the characteristics of Diffusion Transformer \citep{peebles2023dit} and Hourglass transformers \citep{nawrot2022hierarchical}, and employing local attention mechanisms \citep{hassani2023natten}, HDiT offers an efficient model structure suitable for training the diffusion model in the data space.

\begin{figure}[ht]
\centering
\includegraphics[width=0.8\textwidth]{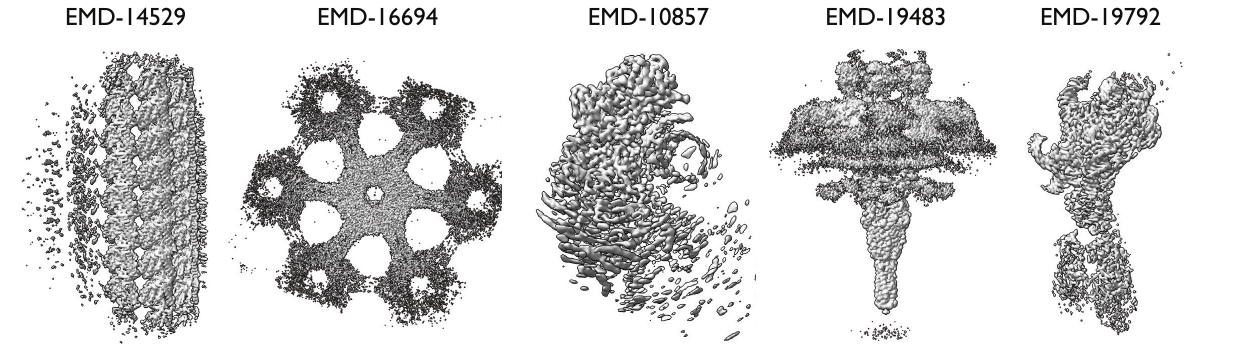}
\caption{\rebuttal{Examples of problematic cases removed from the training dataset after manual curation. Low contour levels are used to highlight the heterogeneous local resolution.}}
\label{fig:supp_data}
\end{figure}

\section{\rebuttal{Additional Data Curation and Processing}}

\subsection{\rebuttal{Data Curation}}
\label{sec:data_filter}
\rebuttal{Examples of the problematic cases which were excluded from the training set after manual curation are shown in \Figref{fig:supp_data}. These includes density maps with very heterogeneous local resolutions that significant portions of the maps are poorly resolved. The EMDB IDs of the training and testing data used in this paper have been uploaded to \href{https://figshare.com/s/9ef2614108391c04d910}{https://figshare.com/s/9ef2614108391c04d910}.}




\subsection{\rebuttal{Data Standardization}}
\label{sec:data_norm}
\rebuttal{We preprocessed the deposited density maps to normalize their values to a consistent range of $[0, 1]$. Since the absolute values of density maps are not intrinsically meaningful and the value ranges vary significantly, we applied a uniform processing procedure as follows: \textbf{Clipping}: We clipped the values of each density map based on the minimum value, which is set to $0$, and the $99.999$th percentile value of the density map. \textbf{Scaling}: Subsequently, the clipped data were scaled to the $[0, 1]$ range by adjusting according to the new minimum and maximum values obtained after clipping. \textbf{Normalization}: we used a mean of $0.04$ and a standard deviation of $0.09$, which were calculated based on the data that have been scaled to the $[0, 1]$ range.}


\section{Additional Detailed Information of \ours}

\subsection{Implementation Details}\label{sec:supp_impl_detail}
\begin{figure}[t]
\centering
\includegraphics[width=\textwidth]{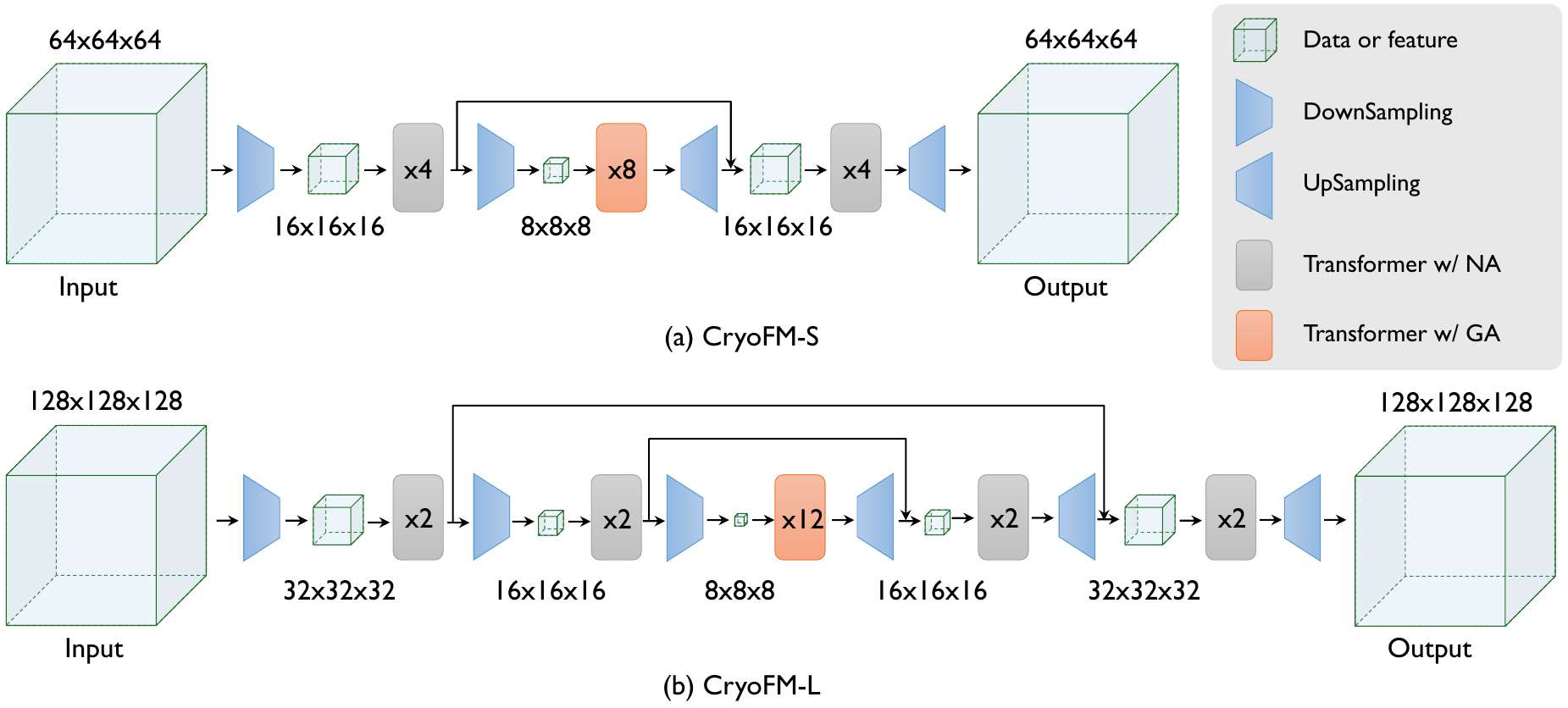}
\caption{Overview of two 3D \ours model details. (a) \ours-S design for input shape $64^3$ and (b) \ours-L for shape $128^3$.}
\label{fig:hdit3d_arch_details}
\end{figure}

\begin{figure}[t]
\centering
\includegraphics[width=\textwidth]{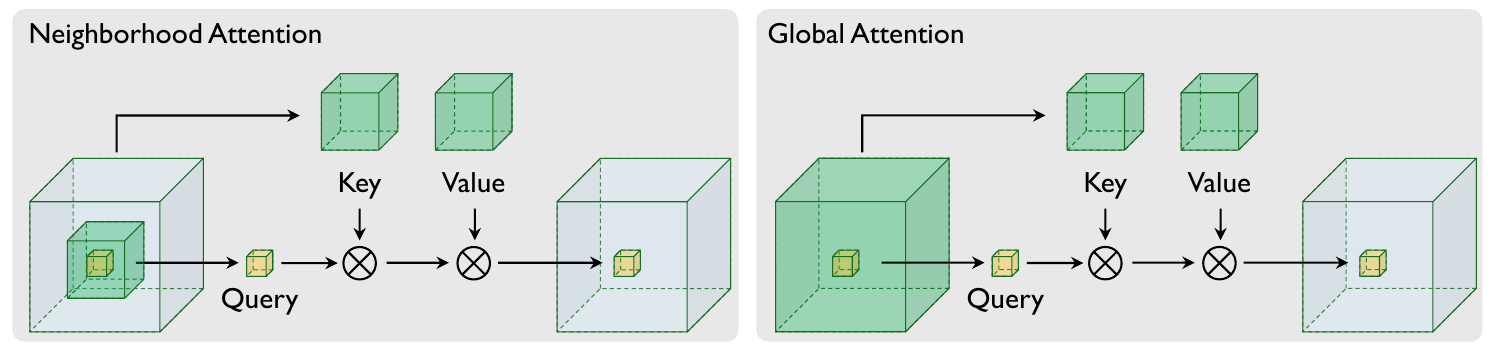}
\caption{Illustration of QKV structure of neighborhood attention (NA) and global attention (GA).}
\label{fig:attn_details}
\end{figure}

\Figref{fig:hdit3d_arch_details} illustrate the model architectures for input dimensions of $64^3$ and $128^3$. \Figref{fig:attn_details} shows the details of different attention structure we used in the transformer block. \Tabref{tab:hdit3d_setup} details the specific model parameter configurations (aligned with the naming conventions used in \citet{crowson2024hdit}) and the training hyperparameters.

In all experiments, we employed the FairseqAdam \citep{ott2019fairseq} optimizer with a default learning rate of 1e-4, betas set to (0.9, 0.98), and a weight decay of 0.01. A linear warm-up strategy was applied during the first 2000 steps of training.

\begin{table}[ht]
\caption{Details of training and model setup.}
\label{tab:hdit3d_setup}
\centering
\begin{tabular}{lcc}
\toprule
\textbf{Parameter}                          & \ours-S & \ours-L \\
\midrule
Parameters                         & 335.18 M              & 308.54 M               \\
GFLOP/forward                      & 395.87                & 427.26                 \\
Training Steps                     & 150k                  & 300k                   \\
Batch Size                         & 128                   & 128                    \\
Precision                          & bf16                  & bf16                   \\
Training Hardware                  & 8$\times$A100                & 8$\times$A100                 \\
\midrule
Patchifying                        & 4                     & 4                      \\
Levels (Local + Global Attention)  & 1 + 1                 & 2 + 1                  \\
Depth                              & {[}4, 8{]}            & {[}2, 2, 12{]}         \\
Widths                             & {[}768, 1536{]}       & {[}320, 640, 1280{]}   \\
Attention Heads (Width / Head Dim) & {[}12, 24{]}          & {[}5, 10, 20{]}        \\
Attention Head Dim                 & 64                    & 64                     \\
Neighborhood Kernel Size           & 7                     & 7                     \\
\bottomrule
\end{tabular}
\end{table}

\subsection{Likelihood Estimation}\label{sec:appendix_nll}

Given a data point $\rvx_0\in\mathbb{R}^n$, we estimate its likelihood by solving a probability flow ODE \citep{song2021scorebased,lipman2022flow,chen2018neural}. We start with the continuity equation of a velocity field $\nnv:[0,1]\times\mathbb{R}^n\to\mathbb{R}^n$:
\begin{align*}
    \frac{d}{dt}\log p_t(\rvx_t) + \nabla \cdot \nnvtxt = 0.
\end{align*}
Integrating $t\in [0,1]$ gives:
\begin{align*}
     \log p_0(\rvx_0) = \log p_1(\rvx_1) - \int_0^1 \nabla \cdot \nnvtxt dt.
\end{align*}
Since $p_1$ is a standard normal distribution, $\log p_1(\rvx_1)$ can be calculated exactly. The second term in the right-hand side of the equation can be computed by solving an ODE:
\begin{align*}
    \frac{d}{dt} \left[ \begin{array}{cc} 
    \rvx_t \\  f_t  
    \end{array} \right]
    =
    \left[ \begin{array}{cc} 
    \nnvtxt \\  \nabla \cdot \nnvtxt 
    \end{array} \right],
\end{align*}
with initial conditions $\rvx_0$ and $f_0=0$. Since the divergence operator requires a lot of computational cost, we follow \citet{grathwohl2018ffjord} to use the Hutchinson trace estimator to get an unbiased estimate:
\begin{align*}
    \frac{d}{dt} \left[ \begin{array}{cc} 
    \rvx_t \\  f_t  
    \end{array} \right]
    =
    \left[ \begin{array}{cc} 
    \nnvtxt \\  \rvz^\top \nabla \nnvtxt \rvz
    \end{array} \right],
\end{align*}
where $\rvz\sim\mathcal{N}(0, \mI)$.
In practice, we use the Runge–Kutta method \citep{dormand1980family} to solve the ODE to get $f_1$, and the log probability is $\log p_0(\rvx_0) = \log p_1(\rvx_1) - f_1$.

\paragraph{Normalization} A commonly employed trick for training flow models is to linearly transform a data point, so that the transformed data are centered around $0$ and the standard deviation is not too off. An additional step is required to estimate the likelihood of the original data point, since the transformation changes the probability density. Given a volume density in the dataset $\rvx\in\mathbb{R}^{64\times 64\times 64}$, if we normalize it by $\rvy=(\rvx - 0.04) / 0.09$ and compute the log-probability of $\log p(\rvy)$, then the probability of $\rvx$ can be further calculated by:
\begin{align*}
    \log p(\rvx) &= \log p(\rvy)+\log \left(\frac{1}{0.09}\right)^{64^3} \\
    &= \log p(\rvy)+ 64^3 \log \left(\frac{1}{0.09}\right) \\
    &\approx \log p(\rvy)+ 631,228
\end{align*}

\section{Additional information of the experiments}

\subsection{Metrics}\label{sec:supp_metrics}

\textbf{$\text{FSC}_\text{AUC}$}~~The area under the curve (AUC) of the FSC curve ($\text{FSC}_{\text{AUC}}$) provides an overall correlation across all spatial frequencies, offering a comprehensive view of the alignment between the ground truth and the restored density map \citep{harauz1986exact}.

\textbf{$\text{FSC}_{0.5}$}~~The FSC resolution at the 0.5 cutoff ($\text{FSC}_{0.5}$) is a standard metric in cryo-EM, indicating the resolution of the reconstructed map relative to the ground truth \citep{rosenthal2003optimal}.

\textbf{Fail Rate (FR)}~~The fail rate is defined as the portion of the test cases where the method either fails to run or produces an $\text{FSC}_{0.5}$ between the result and the ground truth greater than 30 \AA. This metric helps identify cases where the reconstruction process fails to achieve meaningful results, ensuring a more comprehensive evaluation of the method's robustness.

Both $\text{FSC}_{\text{AUC}}$ and $\text{FSC}_{0.5}$ are reported as the mean and standard deviation across the 32 density maps in the test set, excluding failed cases.

\subsection{Spectral Noise Denoising}\label{sec:supp_spectral_noise_results}
The results of spectral noise denoising at additional resolutions are presented in \Tabref{tab:supp_spectral_noise_results}.

\begin{table}[t]
\caption{Additional results of the spectral noise denoising task, comparing \ours with DeepEMhancer \citep{sanchez2021deepemhancer} and EMReady \citep{he2023improvement}.}\label{tab:supp_spectral_noise_results}
\centering
\begin{tabular}{lcccccccccc}
\toprule
{} & \multicolumn{6}{c}{Estimated resolution at different level of noise} \\
\cmidrule{2-7}
& \multicolumn{3}{c}{$8.5$~\AA} & \multicolumn{3}{c}{$15.0$~\AA} \\
\cmidrule(r){2-4} \cmidrule(r){5-7}
{} & FR$\downarrow$ & $\text{FSC}_{\text{AUC}}\uparrow$ & $\text{FSC}_{0.5}\downarrow$ & FR$\downarrow$ & $\text{FSC}_{\text{AUC}}\uparrow$ & $\text{FSC}_{0.5}\downarrow$ \\
\cmidrule(r){1-1} \cmidrule(r){2-4} \cmidrule(r){5-7}

After degradation & - & $0.3275$ & $8.60$ & - & $0.2014$ & $15.98$ \\
\cmidrule(r){1-1} \cmidrule(r){2-4} \cmidrule(r){5-7}

DeepEMhancer & $0.13$ & $0.37\pm0.02$ & $\mathbf{7.4\pm0.3}$ & $0.03$ & $0.16\pm0.04$ & $17.5\pm2.5$ \\

EMReady & $\mathbf{0.00}$ & $0.37\pm0.04$ & $7.7\pm0.8$ & $0.03$ & $0.22\pm0.02$ & $\mathbf{13.3\pm1.6}$ \\

\ours & $\mathbf{0.00}$ & $\mathbf{0.38\pm0.01}$ & $7.5\pm0.3$ & $\mathbf{0.00}$ & $\mathbf{0.24\pm0.01}$ & $15.0\pm0.7$ \\
\bottomrule
\end{tabular}
\end{table}

\section{Estimation of the \rebuttal{forward} operators}\label{sec:degrade_estimation}

\subsection{Spectral noise power estimation}\label{sec:spectral_noise_estimation}

Given two half maps from a cryo-EM reconstruction, $\rvy_1$ and $\rvy_2$ (observations), where the signal and noise power are the same and the signal and noise are uncorrelated, the obeservation models for $\rvy_1$ and $\rvy_2$ are:
\begin{align*}
    \fy_1 = \fV + \rvepsilon_1, \quad \rvepsilon_1(\nu) \sim \mathcal{N}(0, \sigma_{\text{noise}}^2(\nu)), 
\end{align*}
\begin{align*}
    \fy_2 = \fV + \rvepsilon_2, \quad \rvepsilon_2(\nu) \sim \mathcal{N}(0, \sigma_{\text{noise}}^2(\nu)). 
\end{align*}
The so-called ``Gold-Standard'' Fourier Shell Correlation (GSFSC) is the FSC between two half maps, which is defined as:
\begin{align*}
    \text{GSFSC}(\nu) \triangleq \frac{\fy_1(\nu) \cdot \fy_2(\nu)}{\|\fy_1(\nu)\| \cdot \|\fy_2(\nu)\|},
\end{align*}
where $\fy(\nu)$ represents the component of $\fy$ at frequency $\nu$ (i.e., all values on a spherical shell), and $\|\fy(\nu)\|^2$ is the radial power spectrum.

From the GSFSC, the signal-to-noise ratio (SNR) can be computed as \citep{rosenthal2003optimal}:
\begin{align*}
    \text{SNR}(\nu) = \frac{\mathbb{E}[\sigma_{\text{signal}}^2(\nu)]}{\mathbb{E}[\sigma_{\text{noise}}^2(\nu)]} = \frac{\text{GSFSC}(\nu)}{1 - \text{GSFSC}(\nu)}.
\end{align*}

Since $\rvy_1$ and $\rvy_2$ share the same signal and noise at each frequency, and the signal $\fV$ and the noise $\rvepsilon$ are uncorrelated, we have:
\begin{align*}
    \mathbb{E}[\|\fy(\nu)\|^2] = \mathbb{E}[\|\fV(\nu) + \rvepsilon(\nu)\|^2] = \mathbb{E}[\sigma_{\text{signal}}^2(\nu)] + \mathbb{E}[\sigma_{\text{noise}}^2(\nu)].
\end{align*}
Thus, 
\begin{align*}
    \mathbb{E}[\|\fy_1(\nu)\|^2] = \mathbb{E}[\|\fy_2(\nu)\|^2] = \mathbb{E}[\sigma_{\text{signal}}^2(\nu)] + \mathbb{E}[\sigma_{\text{noise}}^2(\nu)].
\end{align*}

Given the known SNR, we can derive:
\begin{align*}
    \mathbb{E}[\sigma_{\text{noise}}^2(\nu)] = \frac{\mathbb{E}[\|\fy_1(\nu)\|^2]}{1 + \text{SNR}(\nu)} = \frac{\mathbb{E}[\|\fy_2(\nu)\|^2]}{1 + \text{SNR}(\nu)},
\end{align*}
and
\begin{align*}
    \mathbb{E}[\sigma_{\text{signal}}^2(\nu)] = \frac{\mathbb{E}[\|\fy_1(\nu)\|^2] \cdot \text{SNR}(\nu)}{1 + \text{SNR}(\nu)} = \frac{\mathbb{E}[\|\fy_2(\nu)\|^2] \cdot \text{SNR}(\nu)}{1 + \text{SNR}(\nu)}.
\end{align*}

\subsection{Anisotropic noise power estimation}\label{sec:aniso_noise_estimation}
Anisotropic noise means that the noise is not uniform across all directions; in certain directions, the noise power is stronger or weaker than in others. In the context of spectral noise, anisotropic noise implies that the variance of the noise depends on the direction in Fourier space. We denote by $\xi = (\phi, \theta)$ the anisotropic angular components on the sphere, and by $\nu$ the frequency component, corresponding to the radial component in spherical coordinates. We assume that the signal power $\sigma^2_{\text{signal}}(\nu)$ is isotropic, meaning $\sigma^2_{\text{signal}}(\xi, \nu)$ is constant for different values of $\xi$.

The noise power averaged over all directions for a given frequency $\nu$ is denoted as:
\begin{align*}
    h(\nu) = \frac{1}{N_\nu} \sum_{\xi} \sigma^2_{\text{noise}}(\xi, \nu),
\end{align*}
where $N_\nu$ represents the number of voxels in the frequency shell at $\nu$.

Our goal is to estimate the anisotropic noise power, $\sigma^2_{\text{noise}}(\xi, \nu)$, in spherical coordinates. We assume that for a given angle $\xi$, the noise power $\sigma^2_{\text{noise}}(\xi, \nu)$ can be expressed as:
\begin{align*}
    \sigma^2_{\text{noise}}(\xi, \nu) = \alpha(\xi, \nu) \cdot \sigma^2_{\text{noise}}(\xi_0, \nu),
\end{align*}
where $\xi_0$ represents an arbitrary reference direction in spherical coordinates, serving as the baseline for the noise power $\sigma^2_{\text{noise}}(\xi, \nu)$, with $\alpha(\xi, \nu)$ adjusting the noise power in other directions. In practice, $\alpha(\xi, \nu)$ can be inferred from the number of particles contributing to each pose during the reconstruction process.

Substituting the anisotropic noise power expression, we can derive:
\begin{align*}
    h(\nu) = \frac{1}{N_\nu} \sum_{\xi} \alpha(\xi, \nu) \cdot \sigma^2_{\text{noise}}(\xi_0, \nu).
\end{align*}
Simplifying, we obtain:
\begin{align*}
    h(\nu) = \frac{\sigma^2_{\text{noise}}(\xi_0, \nu)}{N_\nu} \sum_{\xi} \alpha(\xi, \nu),
\end{align*}
where $h(\nu)$ is the average noise power derived from the the half-map GSFSC in Section \ref{sec:spectral_noise_estimation}.

Thus, we can solve for $\sigma^2_{\text{noise}}(\xi_0, \nu)$ as:
\begin{align*}
    \sigma^2_{\text{noise}}(\xi_0, \nu) = \frac{h(\nu) \cdot N_\nu}{\sum_{\xi} \alpha(\xi, \nu)}.
\end{align*}
Finally, the anisotropic noise power for any angle $\xi$ is given by:
\begin{align*}
    \sigma^2_{\text{noise}}(\xi, \nu) = \alpha(\xi, \nu) \cdot \frac{h(\nu) \cdot N_\nu}{\sum_{\xi} \alpha(\xi, \nu)}.
\end{align*}

\section{Additional Algorithms}
\label{sec:algorithms}

In this section, we present algorithms for addressing different downstream tasks within the flow posterior sampling framework. For spectral noise and anisotropic noise denoising, it is natural to apply \Algref{alg:fps} in the main text since the loss function is well-defined. However, regarding missing wedge restoration and \textit{ab initio} modeling, we can make slight modifications to the algorithm to better solve these problems.

\subsection{Missing Wedge Restoration}

For certain tasks, such as missing wedge restoration, it is not necessary to compute the gradient with respect to the loss function between $\rvy$ and $\mathcal{A}\rvx$, since we can directly attain the optimum in an analytical form. 
The missing wedge restoration is an in-painting task, where a noiseless part of the signal is already observed. $\mathcal{A}$ is an operator extracting a part of the signal, hence the inverse of $\mathcal{A}$ just fills the observed part into the signal. In \Algref{alg:missing_wedge}, we formulate it as the pseudo-inverse operator $\mathcal{A}^\dagger: \mathbb{R}^m \to \mathbb{R}^n$ \citep{kawar2021snips,kawar2022ddrm}.

\begin{algorithm}[t]
\caption{Flow Posterior Sampling for the Missing Wedge Problem}
\label{alg:missing_wedge}
\begin{algorithmic}
\Require a pretrained vector field \rebuttal{$\nnv:[0,1]\times\mathbb{R}^n\to\mathbb{R}^n$}, a \rebuttal{forward} operator $\mathcal{A}:\mathbb{R}^n\to\mathbb{R}^m$, an obseravation $\rvy\in\mathbb{R}^m$, number of steps $N$
\Ensure the recovered signal $\rvx_0$
\State $\rebuttal{\Delta t \gets \frac{1}{N}}$
\State $\rvx_1 \gets \mathcal{N}(0,\mI)$
\For {$t \in [1, 1-\rebuttal{\Delta t}, 1-\rebuttal{2 \Delta t}, \cdots, \rebuttal{\Delta t}]$}
    \State $\hat{\rvx}_0 \gets \rvx_t- t \cdot \rebuttal{\nnvtxt}$ 
    \State $\hat{\rvx}'_0 \gets \mathcal{A}^\dagger\rvy + (1-\mathcal{A}^\dagger)\hat{\rvx}_0$ \Comment{Fill the observed part $\rvy$ into $\hat{\rvx}_0$}
    \State \rebuttal{$\rvx_{t-\Delta t}\gets \frac{t-\Delta t}{t}\cdot \rvx_{t}+\frac{\Delta t}{t} \cdot \hat{\rvx}'_{0}$}
\EndFor
\State \Return $\rvx_0$
\end{algorithmic}
\end{algorithm}

\subsection{\textit{Ab initio} Modeling}

For \textit{ab initio} modeling, the $k$-th \rebuttal{observed} volume (we will name it the \textit{projection} later) is $\mathcal{A}^{(k)}(\mV) = \mathcal{P}(\phi^{(k)}, \mV)$, where $\phi^{(k)}\in\mathbb{SO}(3)\times\mathbb{R}^2$ is the pose.
The posterior sampling of \textit{ab initio} modeling is more complex than the previous tasks in that the pose $\phi^{(k)}$ can not be easily pre-determined. The setting is closely related to blind diffusion posterior sampling \citep{chung2023blind,kapon2024mas,laroche2024fast,gan2024block}, where the \rebuttal{forward} operator is unknown. We find $\phi^{(k)}$ in the sampling process, by minimizing the discrepancy between the projection and the observation $\vy^{(k)}$ using a coarse-to-fine strategy \citep{zhong2021cryodrgn2}. The discrepancy is defined by the negative correlation between a projection $\rvp$ and an observation $\rvy$:
\begin{align*}
    \psi_\mathrm{cor}(\rvp, \rvy)=-\frac{\rvp \cdot \rvy}{ \|\rvp\| \cdot \|\rvy\|}.
\end{align*}
When computing the likelihood, we further introduce two variants of the loss function to make the sampling process more robust, since this setting is highly ill-posed compared to others. The first one applies a stop gradient operator (denoted by ``$\mathrm{sg}$'') to the norm of the projection, where the stop gradient operator is identity at forward computation time but has zero partial derivatives \citep{van2017neural,roy2018theory,jang2016categorical}:
\begin{align*}
    \psi_\mathrm{sg}(\rvp, \rvy)=-\frac{\rvp \cdot \rvy}{ \mathrm{sg}(\|\rvp\| \cdot \|\rvy\|)}.
\end{align*}
This variant does not constrain the norm of the projection, as long as it has a large correlation with the observation. The second one penalizes additional noises in the projection so that it has a value of $0$ in the background area of the observation $\rvy$. The background area is defined by a set of indices $\mathcal{B}(q, \rvy)$ where the corresponding value is smaller the $q$-th percentile of $\rvy$.
\begin{align*}
    \psi_{\mathrm{bg}(q)}(\rvp, \rvy )=\sum_{i \in \mathcal{B}(q, \rvy)} \rvp_i^2.
\end{align*}
The final loss function $\mathcal{L}$ can be a combination of $\psi_\mathrm{cor}$, $\psi_\mathrm{sg}$ and $\psi_{\mathrm{bg}(q)}$. We tune the parameter on different datasets (\Tabref{tab:abinitio}) by examining the human preference for the reconstruction results. A full alorithm can be found in \Algref{alg:ab_initio}.

\begin{algorithm}[t]
\caption{Flow Posterior Sampling for \textit{Ab-initio} Modeling}
\label{alg:ab_initio}
\begin{algorithmic}
\Require a pretrained vector field \rebuttal{$\nnv:[0,1]\times\mathbb{R}^n\to\mathbb{R}^n$}, $K$ obseravations $\rvy^{(k)}\in\mathbb{R}^m$, number of steps $N$, maximum step size $\lambda_{\text{max}}$ at each step, the loss function with respect to the likelihood term $\mathcal{L}$
\Ensure the recovered signal $\rvx_0$
\State $\rebuttal{\Delta t \gets \frac{1}{N}}$
\State $\rvx_1 \gets \mathcal{N}(0,\mI)$
\For {$t \in [1, 1-\rebuttal{\Delta t}, 1-\rebuttal{2 \Delta t}, \cdots, \rebuttal{\Delta t}]$}
    \State $\rvx'_{t-\rebuttal{\Delta t}} \gets \rvx_t-\rebuttal{\Delta t \cdot \nnvtxt}$ 
    \For {$k \in [1, 2, \cdots, K]$}
        \State $\phi^{(k)} \gets \underset{\phi\in\mathbb{SO}(3)\times\mathbb{R}^2}{\arg\min}~\psi_\mathrm{cor} \left(\mathcal{P}(\phi, \hat{\rvx}_0(\rvx_t)), \rvy^{(k)}\right)$ \Comment{Search the optimal pose}
        \State $l^{(k)}(\rvx_t) \gets  \mathcal{L}\left(\mathcal{P}(\phi^{(k)}, \hat{\rvx}_0(\rvx_t)), \rvy^{(k)}\right)$ 
        \Comment{Loss w.r.t the optimal pose}
    \EndFor
    \State $l(\rvx_t)\gets\frac{1}{K}\sum_k l^{(k)}(\rvx_t)$
    \State $\rvg \gets \frac{\nabla_{\rvx_{t}} l(\rvx_t)}{\|\nabla_{\rvx_{t} } l(\rvx_t) \|_2}$ 
    \State $\lambda_t \gets \min\left\{\lambda_{\text{max}}, \frac{t}{1-t}\right\}$ 
    \State $\rvx_{t-\rebuttal{\Delta t}} \gets \rvx'_{t-\rebuttal{\Delta t}} \rebuttal{-} \lambda_t \rvg$ 
\EndFor
\State \Return $\rvx_0$
\end{algorithmic}
\end{algorithm}

\begin{table}[h]
    \caption{Hyperparameters for \textit{ab initio} modeling}
    \label{tab:abinitio}
    \centering
    \begin{tabular}{cccc}
        \toprule
        EMPIAR-ID & stop gradient & $q$ \\
        \midrule 
        $10345$ (synthetic) & no & $0.7$ \\ 
        $10028$ & yes & $0.0$ \\ 
        $10827$ & no & $0.7$ \\ 
        \bottomrule
    \end{tabular}
\end{table}

\section{Additional Ablation Study}
\label{sec:supp_ablate}

\subsection{Data Augmentation} \label{sec:ablate_data}
\begin{figure}[t]
\centering
\includegraphics[width=\textwidth]{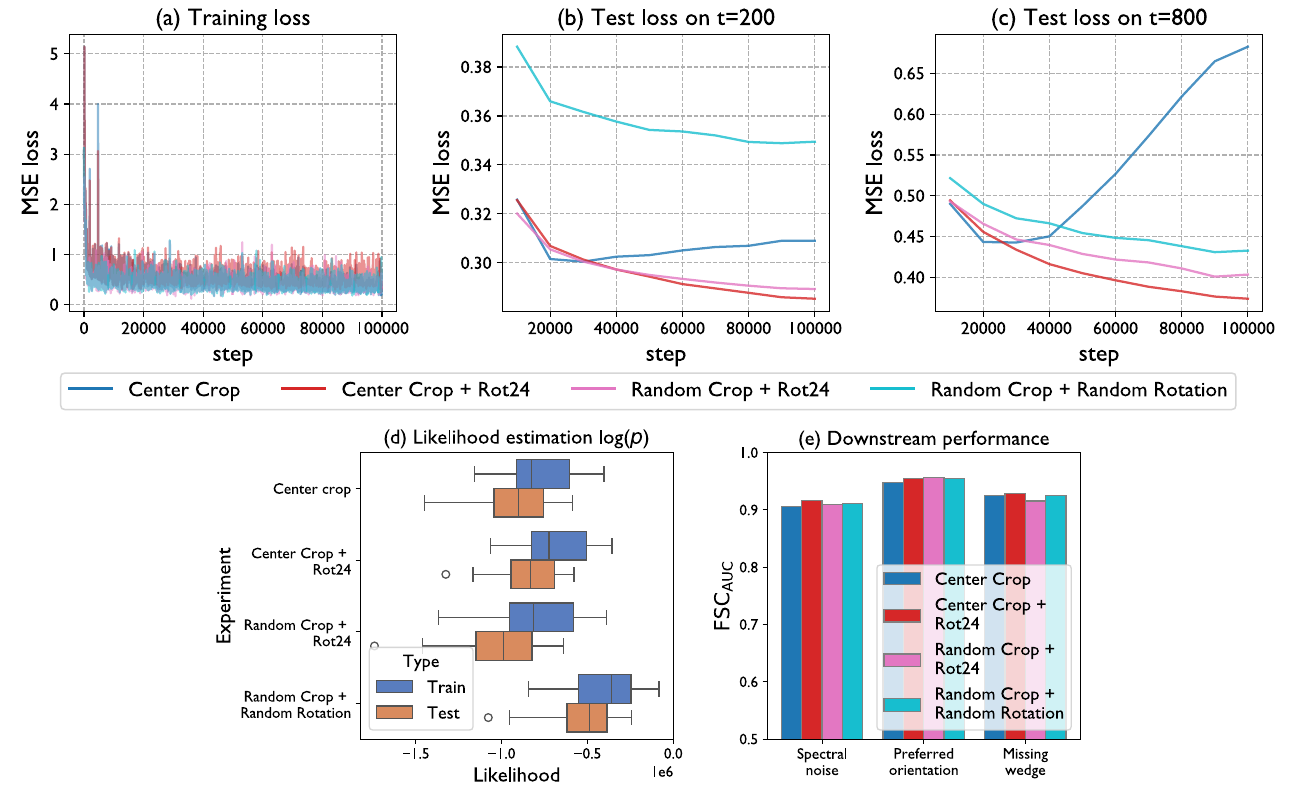}
\caption{Results on different data augmentation strategies.}
\label{fig:ablate_data_aug}
\end{figure}

Protein density data are three-dimensional data characterized by large size differences and lack of canonical orientation. To train the model efficiently, we need to crop the data to the same size. Additionally, to expose the model to enough data with various orientations, we also need to apply appropriate rotational augmentation to the data. In this section, we discuss different data augmentation strategies and how they affect the training and the performance on the downstream tasks.
We implemente four different augmentation strategies, as shown in \Figref{fig:ablate_data_aug}:
\textbf{center crop}: Only crops a patch from the center of the original data for training;
\textbf{random crop}: Randomly crops patches from the original data;
\textbf{rot24}: Achieves data rotation by swapping axes order and flipping the data (resulting in 24 possible orientations);
\textbf{random rotation}: Randomly rotates data in the $\mathrm{SO}(3)$ space.

As seen in \Figref{fig:ablate_data_aug} (a), all strategies enabled the model's loss to converge stably on the training set. However, \Figref{fig:ablate_data_aug} (b) and (c) indicate that the weakest augmentation strategy, center crop, leads to some overfitting on the test set. From the estimated likelihood of the data in the training and test set in \Figref{fig:ablate_data_aug} (d), model exhibits similar behavior across different strategies. From the downstream task performance shown in \Figref{fig:ablate_data_aug} (e), the differences among the strategies were negligible. Given that our modeling targets a more holistic $p(\rvx)$, we choose random crop combined with random rotation as our augmentation strategy.

\subsection{Patchifying \& Downsampling Factor} \label{sec:ablate_patchify_downsample}

For three-dimensional data, the number of tokens is proportional to the cube of the edge length $D$ of the data. Applying patchification to the input and down-sampling within the model significantly impacts the number of tokens the transformer block needs to process, thereby influencing the model's efficiency. In this section, we conducted ablation experiments on the number of patchifying and down-sampling operations. As shown in \Tabref{tab:ablate_model_arch}, the default configuration we used is p4-d1, where p4 represents a patchifying factor of 4, meaning a $4\times4\times4$ patch in the input is flattened and used as a token vector, and d1 represents that the data is down-sampled once in the model. We compare configurations with p $= 2, 4, 8$ and d = $0, 1, 2$.

\begin{figure}[t]
\centering
\includegraphics[width=0.85\textwidth]{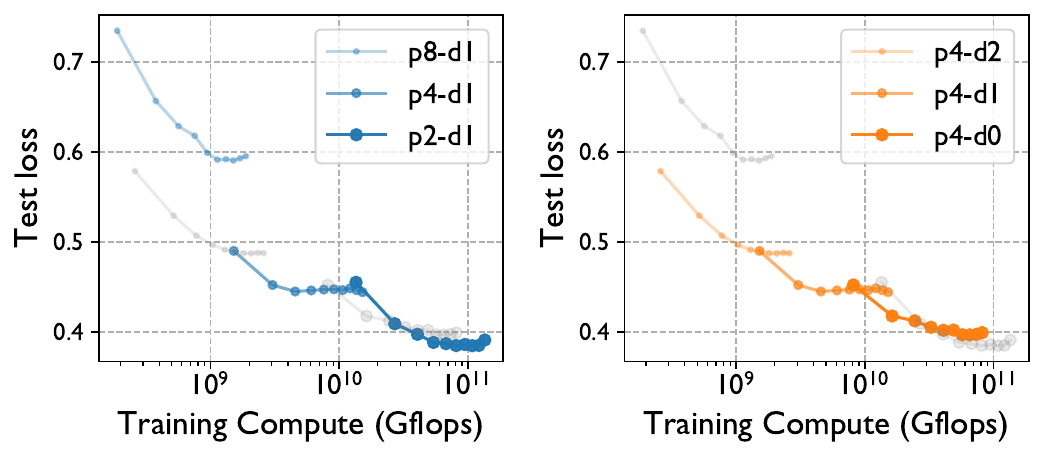}
\caption{MSE loss on test set as a function of total training compute.}
\label{fig:ablate_model_arch_appendix}
\end{figure}

In \Figref{fig:ablate_model_arch_appendix}, we present the loss curves on the test set during training for different model variants. Given the substantial differences in the computational costs among these model variants, the horizontal axis represents the training compute in Gflops. The estimation of training compute is: model Gflops $\cdot$ batch size $\cdot$ training steps $\cdot$ 3 \citep{peebles2023dit}. We observe that the models with higher computational costs achieve lower loss levels, indicating the potential for scalability. On the other hand, as shown in \Tabref{tab:ablate_model_arch}, the inference speeds of the p2 and d0 models are exceedingly slow. Therefore, we selected the p4-d1 configuration with a comprehensive consideration of model performance and efficiency in downstream applications.

\begin{table}[t]
\caption{Different model configurations.}
\label{tab:ablate_model_arch}
\centering
\begin{tabular}{lccccc}
\toprule
 & Default & \multicolumn{2}{c}{Patchifying variants}  & \multicolumn{2}{c}{Down-sampling variants} \\
\cmidrule(l{2pt}r{2pt}){2-2} \cmidrule(l{2pt}r{2pt}){3-4} \cmidrule(l{2pt}r{2pt}){5-6}
Short name & p4-d1          & p2-d1              & p8-d1          & p4-d0                & p4-d2               \\
\cmidrule(l{2pt}r{2pt}){1-1} \cmidrule(l{2pt}r{2pt}){2-2} \cmidrule(l{2pt}r{2pt}){3-4} \cmidrule(l{2pt}r{2pt}){5-6}
Patching   & 4               & 2                   & 8               & 4                     & 4                    \\
Depth      & {[}4, 8{]}      & {[}4, 8{]}          & {[}4, 8{]}      & {[}12{]}              & {[}2, 2, 8{]}        \\
Width      & {[}768, 1536{]} & {[}768, 1536{]}     & {[}768, 1536{]} & {[}1536{]}            & {[}384, 768, 1536{]} \\
Parameters  & 335.18 M        & 335.10 M            & 335.87 M        & 377.75 M              & 316.06 M            \\ 
Thru. (FPS) & 10.23 & 1.19 & 40.61 & 2.67 & 28.52 \\
\bottomrule
\end{tabular}
\end{table}

\subsection{Hyperparameters for Posterior Sampling} \label{sec:ablate_dps_hyp}

Posterior sampling process involves continuously moving forward and correcting in the directions of prior $ p(\rvx) $ and the likelihood $ p(\rvx|\rvy) $. In this section, we present ablation experiments on the two parameters that significantly impact the effectiveness of the posterior sampling algorithm: the number of total timesteps, and $\lambda_\text{max}$ which controls the extent to which $\rvx_t$ is modified based on the observed data. \rebuttal{We evaluated the performance using only the first four samples from the test set.}


\Tabref{tab:ablate_dps_hyp_spectral} and \Tabref{tab:ablate_dps_hyp_missing} present the ablation studies on these two parameters across three downstream tasks. The results show that for a fixed $\lambda_\text{max}$, increasing the number of sampling timesteps improves quality, with the best performance at $1000$ timesteps. Moreover, a $\lambda_\text{max}$ of $5$ yields optimal or near-optimal results. Therefore, we defaulted to a $\lambda_\text{max}$ of $5$ and $1000$ timesteps for the posterior sampling parameters in our main results.

\begin{table}[t]
\caption{Results with different posterior sampling parameters on spectral noise and anisotropic noise denoising task.}
\label{tab:ablate_dps_hyp_spectral}
\centering

\resizebox{\textwidth}{!}{
\begin{tabular}{cccccccccc}
\toprule
{} & {} & \multicolumn{5}{c}{Spectral Noise Denoising} & \multicolumn{3}{c}{Anistropic Noise Denoising} \\
{} & {} & $3.2$~\AA & $4.3$~\AA & $6.1$~\AA & $8.5$~\AA & $15.0$~\AA & $\pm45^\circ$ & $\pm30^\circ$ & $\pm15^\circ$ \\
\cmidrule(l{2pt}r{2pt}){3-7} \cmidrule(l{2pt}r{2pt}){8-10}
\multicolumn{2}{c}{After degradation} & $0.8874$ & $0.5526$ & $0.4968$ & $0.3278$ & $0.2000$ & $0.6626$ & $0.6326$ & $0.6113$ \\
\cmidrule(l{2pt}r{2pt}){1-2} \cmidrule(l{2pt}r{2pt}){3-7} \cmidrule(l{2pt}r{2pt}){8-10}
$\lambda_{\text{max}}$ & \# timesteps & & & & & & & & \\
$1.0$ & $50$ & $0.2511$ & $0.2573$ & $0.1811$ & $0.1736$ & $0.1349$ & $0.1994$          & $0.1638$          & $0.1391$          \\
$1.0$ & $200$ & $0.7301$ & $0.5525$ & $0.4985$ & $0.3855$ & $0.2476$ & $0.5965$          & $0.4836$          & $0.4126$          \\
$1.0$ & $1000$ & $0.9512$ & $0.6771$ & $0.6082$ & $0.3928$ & $0.2460$ & $0.8745$          & $0.8191$          & $0.7741$          \\
$5.0$ & $50$ & $0.4618$ & $0.4140$ & $0.3577$ & $0.3140$ & $0.2156$ & $0.3778$          & $0.8191$          & $0.2576$          \\
$5.0$ & $200$ & $0.8994$ & $0.6417$ & $0.5808$ & $\mathbf{0.4115}$ & $\mathbf{0.2678}$ & $0.7923$          & $0.6694$         & $0.5714$         \\
$5.0$ & $1000$ & $\mathbf{0.9545}$ & $\mathbf{0.6843}$ & $\mathbf{0.6191}$ & $0.3794$ & $0.2313$ & $\mathbf{0.8879}$ & $\mathbf{0.8453}$ & $\mathbf{0.8053}$ \\
$10.0$ & $50$ & $0.4989$ & $0.4312$ & $0.3904$ & $0.3381$ & $0.2249$ & $0.4012$          & $0.3027$          & $0.2568$          \\
$10.0$ & $200$ & $0.9025$ & $0.6397$ & $0.5798$ & $\underline{0.4103}$ & $\underline{0.2631}$ & $0.7926$          & $0.6713$          & $0.5503$          \\
$10.0$ & $1000$ & $\underline{0.9536}$ & $\underline{0.6649}$ & $\underline{0.6183}$ & $0.3752$ & $0.2314$ & $\underline{0.8780}$    & $\underline{0.8329}$    & $\underline{0.7926}$   \\
\bottomrule
\end{tabular}
}

\end{table}

\begin{table}[ht]
\caption{Results with different sampling steps on missing wedge restoration task.}
\label{tab:ablate_dps_hyp_missing}
\centering
\begin{tabular}{lc}
\toprule
                  & $\text{FSC}_{\text{AUC}}\uparrow$ \\
\cmidrule{2-2}
After degradation & $0.7988$ \\
\cmidrule(l{2pt}r{2pt}){1-1} \cmidrule(l{2pt}r{2pt}){2-2}
Timesteps = $50$ & $0.9295$ \\
Timesteps = $200$ & $0.9295$ \\
Timesteps = $1000$ & $\mathbf{0.9310}$ \\
\bottomrule
\end{tabular}
\end{table}

\subsection{Latent Flow Models}\label{sec:ablate_lfm}

In this section, we discuss the choice of the input space of the flow-matching model: should we learn in the \textit{voxel space} or the \textit{latent space}? Most of the early works \citep{goodfellow2014generative,van2016pixel,brock2018large,dhariwal2021diffusion} for generative modeling learned the data distribution directly on the original image space.  \citet{rombach2022high} pointed out that learning diffusion models on the latent space can generate high-quality images with much less computational cost. The key is to learn an variational auto-encoder (VAE) \citep{kingma2013auto} which provides a low-dimensional representation that is equivalent to the original data space \citep{rombach2022high,podell2023sdxl}. Then, a Diffusion Transformer (DiT) \citep{peebles2023dit} is trained on the latent codes of the VAE.

We have trained several latent flow models (LFMs). An LFM-$d$ model is comprised of a VAE-$d$ and a DiT, where $d$ denotes the downsampling factor of the VAE. A larger downsampling factor $d$ leads to greater information compression and increased information loss in the VAE, while it may ease the training of the DiT with smaller size of latent codes.

\Figref{fig:ldm_results} shows the results of unconditional sampling from two latent flow models (LFM): LFM-$4$ and LFM-$8$, where the high-frequency information contains more noise, and secondary structures such as $\alpha$-helicies can not be observed. From the preliminary result, we concluded that: \textit{For protein density modeling, the flow models trained in the voxel space converge much faster and better than the latent flow models.}

Our observation concides with that of \citet{crowson2024hdit}, where they found that latent diffusion models may fail to generate fine details. We attribute the failure of latent flow models to the weak representational power of the learned variational auto-encoder (VAE) \citep{kingma2013auto}. The protein density dataset for training the VAE is much smaller than the large-scale dataset in image synthesis tasks \citep{rombach2022high,podell2023sdxl}. The low-dimensional space of the VAE is poorly regularized and can be sensitive to perturbations. A small error in the latent space may lead to large peceptual loss in the voxel space, especially for cryo-EM tasks which is highly sensitive to fine details. 


\subsubsection{Architecture of LFM}
It is somewhat difficult to conduct a rigorous comparison between latent-space models and voxel-space models as their architectures may differ in several aspects. For the completeness of this work, we will present the results of these models. However, for the sake of simplicity in presentation, we only show two representative settings, and the conclusion is generally consistent across various settings (such as parameter size, model architecture, batch size, etc.).

The procedure of learning a latent flow model whose downsampling factor is $d$ (LFM-$d$) can be divided into two parts:
\begin{itemize}
    \item a VAE-$d$ which compresses a protein density $\rmV\in\mathbb{R}^{D\times D\times D}$ to a latent code $\rvz\in\mathbb{R}^{\frac{D}{d}\times \frac{D}{d}\times \frac{D}{d}}$, $d$ is the compressing factor. The compressing factor governs the balance between perceptual compression and distortion \citep{blau2018perception,rombach2022high}. We employed the publicly available code of Stable Diffusion \footnote{https://github.com/CompVis/stable-diffusion} by replacing 2D convolution layers with 3D layers. We conducted experiments with two settings: $d = 4$ and $d = 8$. 
    \item a DiT (Diffusion Transformer) \citep{peebles2023dit} that learns the distribution of the latent codes. We experimented with a standard DiT architecture by replacing the 2D patchifying module with a 3D module \footnote{https://github.com/huggingface/diffusers}. The DiT has $12$ layers, each of which has $1024$ dimensions, and the patchify factor is set to $2$.
\end{itemize}

Section \ref{sec:appendix_vae_eval} shows that VAE-$8$ performs significantly worse in reconstrution tasks since it is over-compressed and causes loss of information. However, a larger downsampling factor produces smaller number of latent codes, which may ease the training of DiT. Therefore, we trained two DiTs on the latent codes of two VAEs. For each setting, we trained the DiT for $300$k steps.

\subsubsection{Evaluation of VAE} \label{sec:appendix_vae_eval}

Given a volume $\rmV \in \mathbb{R}^{D\times D\times D}$, we evaluate the reconstruction quality of VAE through auto-encoding Fourier shell correlation (AE-FSC) and auto-encoding L-1 loss between the input density and the reconstruction, defined by:
\begin{align*}
    \text{AE-FSC} & \triangleq \text{FSC}(\rmV, \text{VAE}(\rmV)) \\
    \text{AE-L1} & \triangleq \|\rmV - \text{VAE}(\rmV)\|_1 \\
\end{align*}

\Tabref{tab:vae_metric} shows that VAE-$4$ outperforms VAE-$8$ in terms of reconstruction quality, since VAE-$8$ learns to compress a grid of $8\times8\times8=512$ voxels into a latent code, which is $8$ times harder than VAE-$4$ that operates on $4\times4\times4=64$ voxels.

\begin{table}[ht]
    \caption{Reconstruction quality of VAE}\label{tab:vae_metric}
    \centering
    \begin{tabular}{ccccc}
        \toprule
        Model & AE-FSC ($\uparrow$) & AE-L1 ($\downarrow$) & Training Steps\\
        \midrule 
        VAE-$4$ & $0.9821$ & $0.0028$ & $588,480$ \\
        VAE-$8$ & $0.8537$ & $0.0049$ & $1,097,872$ \\
        \bottomrule
    \end{tabular}
\end{table}


\end{document}